\documentclass[aps,prd,twocolumn,groupedaddress,floatfix,showpacs,preprintnumbers]{revtex4}

\usepackage{graphicx}

\newcommand{\lwig}{\mbox{\,\raisebox{.3ex}
    {$<$}$\!\!\!\!\!$\raisebox{-.9ex}{$\sim$}\,}}
\newcommand{\gwig}{\mbox{\,\raisebox{.3ex}
    {$>$}$\!\!\!\!\!$\raisebox{-.9ex}{$\sim$}}\,}

\newcommand{\massnophhalo}{$m_\nu=3.46^{+1.73(4.03)}_{-1.34(2.32)}$~eV }
\newcommand{\massnopheg}{$m_\nu=0.20^{+0.19(0.61)}_{-0.12(0.18)}$~eV, }
\newcommand{\massnophno}{$m_\nu=0.40^{+0.32(0.87)}_{-0.16(0.27)}$~eV, }
\newcommand{\lowbdnopheg}{$m_\nu > 0.02$~eV } 
\newcommand{\chiminnophhalo}{$\chi^2_{\rm min}=15.64$ } 
\newcommand{\chiminnopheg}{$\chi^2_{\rm min}=25.82$ } 
\newcommand{\chiminnophno}{$\chi^2_{\rm min}=8.03$, }
\newcommand{\chiminnozhalo}{$\chi^2_{\rm min}=41.37$}
\newcommand{\chiminnozeg}{$\chi^2_{\rm min}=38.25$}
\newcommand{\massleehalo}{$m_\nu=3.71^{+1.40(3.27)}_{-1.12(1.96)}$~eV, }
\newcommand{\massleeeg}{$m_\nu =0.77^{+0.48(1.36)}_{-0.30(0.51)}$~eV, }

\newcommand{\massrangehaloonesig}{$2.1$~eV~$\leq~m_\nu\leq~6.7$~eV }
 
\newcommand{\massrangeegonesig}{$0.08$~eV~$\leq~m_\nu\leq~1.3$~eV} 
\newcommand{\massrangenoonesig}{$0.24$~eV~$\leq~m_\nu\leq~2.6$~eV} 

\newcommand{\massrangeegstrongonesig}{$0.08$~eV~$\leq~m_\nu\leq~0.40$~eV} 
\newcommand{\massnophegbestfit}{$m_\nu=0.20$~eV}

\begin{document}

\preprint{ITP-Budapest 581\hspace{6ex} DESY 02-014}

\title{Relic neutrino masses and the highest energy cosmic rays}


\author{Z.~Fodor}
\email[]{fodor@pms2.elte.hu}
\affiliation{Institute for Theoretical Physics, E\"otv\"os University, 
P\'azm\'any
1, H-1117 Budapest, Hungary}
\author{S.D.~Katz}
\email[]{sandor.katz@desy.de}
\author{A. Ringwald}
\email[]{andreas.ringwald@desy.de}
\affiliation{Deutsches Elektronen-Synchrotron DESY, Notkestr. 85, D-22607
Hamburg, Germany}



\begin{abstract}
We consider the possibility that a large fraction of the 
ultrahigh energy cosmic rays are decay products of Z bosons 
which were produced in the scattering of ultrahigh energy cosmic neutrinos on cosmological relic neutrinos. 
We compare the observed ultrahigh energy cosmic ray spectrum with the one predicted in the 
above Z-burst scenario and determine the required mass of the heaviest relic neutrino as well as the necessary 
ultrahigh energy cosmic neutrino flux via a maximum 
likelihood analysis. We show that the value of the neutrino mass obtained in this way
is fairly robust against variations in presently unknown quantities,  like
the amount of neutrino clustering, the universal radio background, and the
extragalactic magnetic field, within their anticipated uncertainties. 
Much stronger systematics arises from different  possible assumptions about the diffuse
background of ordinary cosmic rays from unresolved astrophysical sources. 
In the most plausible case that these ordinary cosmic rays are 
protons of extragalactic origin, one is lead to a required neutrino mass in the range
\massrangeegonesig\ at the 68\,\% confidence level.
This range narrows down considerably if a particular universal radio background is assumed, e.g. 
to \massrangeegstrongonesig\ for a large one. 
The required flux of ultrahigh energy cosmic neutrinos near the resonant energy should be detected in the near future 
by AMANDA, RICE, and the Pierre Auger Observatory, otherwise the Z-burst scenario will be ruled out. 
\end{abstract}


\maketitle

\section{Introduction}

Big bang cosmology predicts the existence of a background gas of
free photons and neutrinos. The measured cosmic microwave background (CMB)
radiation supports the applicability of standard cosmology
back to photon decoupling which occured approximately one hundred thousand years
after the big bang. The relic neutrinos, on the other hand,
have decoupled when the universe had a temperature of one MeV and an age
of just one second. Thus, a measurement of the relic neutrinos, with a
predicted average number density of 
\begin{equation}
\label{standard_number_dens}
\langle n_{\nu_i}\rangle_0 = \langle n_{\bar\nu_i}\rangle_0
= \frac{3}{22}\, 
\underbrace{\langle n_{\gamma}\rangle_0}_{\rm CMB}
\simeq 56\ {\rm cm}^{-3}\,,
\end{equation}
per light ($m_{\nu_i}\ll 1$~MeV) neutrino species $i$, 
would provide a new window to the early universe. Their predicted number 
density is comparable to the one of the microwave photons. However,
since neutrinos interact only weakly, the relic neutrinos have not yet been detected directly
in laboratory 
experiments~\cite{Stodolsky:1975aq,Smith:1983jj,Shvartsman:1982sn,Langacker:1983ih,Ferreras:1995wf,Hagmann:1999kf,Duda:2001hd}.

\begin{figure}[b]
\vspace*{2.0mm} 
\includegraphics[width=5.0cm]
{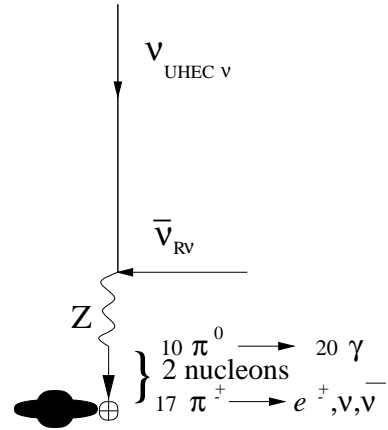}
\caption[...]{\label{illu}
Illustration of a Z-burst resulting from the resonant annihilation of an 
ultrahigh energy cosmic neutrino on a relic (anti-)neutrino.  
}
\end{figure}

Recently, an indirect detection possibility for relic neutrinos has been discussed~\cite{Fargion:1999ft,Weiler:1999sh}.
It is based on so-called Z-bursts resulting from the resonant annihilation of ultrahigh energy cosmic neutrinos
(UHEC$\nu$s) with relic neutrinos into $Z$ bo\-sons~\cite{Weiler:1982qy,Roulet:1993pz,Yoshida:1997ie} 
(cf. Fig.~\ref{illu}). On resonance, the corresponding cross section is enhanced
by several orders of magnitudes. 
If neutrinos have non-vanishing masses $m_{\nu_i}$
-- for which there is ra\-ther convincing evidence in view of the apparent observation of 
neutrino 
oscillations~\cite{Fukuda:1998mi,Fukuda:1998tw,Fukuda:1998ub,Fukuda:2000np,Ambrosio:2001je,Cleveland:1998nv,%
Hampel:1999xg,Abdurashitov:1999zd,Fukuda:2001nj,Fukuda:2001nk,%
Altmann:2000ft,Ahmad:2001an,Athanassopoulos:1995iw,Athanassopoulos:1996wc,Athanassopoulos:1996jb,%
Athanassopoulos:1998pv,Athanassopoulos:1998er,Aguilar:2001ty,Declais:1995su,Apollonio:1999ae,Boehm:2000gk,%
Eitel:2000by,Armbruster:2002mp,Avvakumov:2002jj}
 --
the respective resonance 
energies, in the rest system of the relic neutrinos, correspond to 
\begin{eqnarray}
\label{eres}
E_{\nu_i}^{\rm res} = \frac{M_Z^2}{2\,m_{\nu_i}} = 4.2\cdot 10^{21}\ {\rm eV}  
\left( \frac{1\ {\rm eV}}{m_{\nu_i}}\right)
\,,
\end{eqnarray}
with $M_Z$ denoting the mass of the Z boson. 
These resonance energies are, for neutrino masses of ${\mathcal O}(1)$ eV, remarkably
close to the energies of the highest energy cosmic rays observed at 
Earth by collaborations such as AGASA~\cite{Takeda:1998ps}, Fly's Eye~\cite{Bird:yi,Bird:wp,Bird:1994uy}, 
Haverah Park~\cite{Lawrence:cc,Ave:2000nd}, HiReS~\cite{Kieda00}, and Yakutsk~\cite{Efimov91} (for a review, see 
Ref.~\cite{Nagano:2000ve}). 
Indeed, it was argued~\cite{Fargion:1999ft,Weiler:1999sh}
that the ultrahigh energy cosmic rays (UHECRs) above the  
predicted Greisen-Zatsepin-Kuzmin (GZK) 
cutoff~\cite{Greisen:1966jv,Zatsepin:1966jv} ar\-ound $4\cdot 10^{19}$~eV
are mainly protons (and, maybe, photons) from Z decay. 
In this way, one  possibly also solves one of the outstanding problems of ultrahigh energy cosmic ray
physics~\cite{Bhattacharjee:2000qc}, namely the apparent observation of cosmic rays with energies above the GZK 
cutoff, in an elegant and economical way without 
invoking new physics beyond the Standard Model, except for neutrino masses.  
The GZK puzzle hinges on the fact that
nucleons with super-GZK energies have a short attenuation length of about $50$~Mpc, due
to inelastic interactions with the cosmic microwave background, while plausible astrophysical
sources for those energetic particles are much farther away~\cite{Sigl:1994eg,Elbert:1995zv}. 
Ultrahigh energy neutrinos produced at cosmological distances, on the other hand, can reach the
GZK zone unattenuated and their resonant annihilation on the relic neutrinos could just result in the observed cosmic rays
of the highest energies. Moreover, at present this annihilation process may be the only way to detect 
the relic neutrino background, a basic ingredient of our cosmological picture. 
 
The Z-burst hypothesis for the ultrahigh energy cosmic rays was discussed in many  
papers~\cite{Waxman:1998yh,Yoshida:1998it,Blanco-Pillado:2000yb,Gelmini:1999qa,Gelmini:2000ds,%
Weiler:1999ny,Crooks:2001jw,Gelmini:2000bn,Pas:2001nd,Fargion:2000pv,%
Fodor:2001qy,Fodor:2001rg,McKellar:2001hk,Ringwald:2001mx,Kalashev:2001sh,Gelmini:2002xy}.
The required UHEC$\nu$ fluxes were estimated in Ref.~\cite{Waxman:1998yh} for different spectral
indices. 
In Ref.~\cite{Yoshida:1998it}, particle spectra were determined numerically for case 
studies which supported the Z-burst scenario. 
The effect of possible lepton asymmetries was studied in Ref.~\cite{Gelmini:1999qa}.
In Ref.~\cite{Pas:2001nd}, the analysis of the Z-burst mechanism was advocated as 
one of the few possibilities for an absolute neutrino mass determination.

In the present paper, we present the details of (and extend) our recent 
quantitative investigation of the Z-burst scenario~\cite{Fodor:2001qy} (see also Refs.~\cite{Fodor:2001rg,Ringwald:2001mx}), 
where we have determined the required mass of the heaviest relic neutrino as well as the 
necessary ultrahigh energy cosmic neutrino flux via a maximum 
likelihood analysis. 
But before we start this enterprise, we shall briefly review the current knowledge of
the absolute scale of neutrino masses. 

Neutrinos almost certainly have non-vanishing masses and mix. This follows 
from the apparent observation of neutrino oscillations whose evidence is compelling 
for atmospheric neutrinos~\cite{Fukuda:1998mi,Fukuda:1998tw,Fukuda:1998ub,Fukuda:2000np,Ambrosio:2001je}, 
strong for solar 
neutrinos~\cite{Cleveland:1998nv,Hampel:1999xg,Abdurashitov:1999zd,Fukuda:2001nj,Fukuda:2001nk,%
Altmann:2000ft,Ahmad:2001an}, 
and so-far unconfirmed for neutrinos produced in the laboratory and studied by the LSND 
collaboration and others~\cite{Athanassopoulos:1995iw,Athanassopoulos:1996wc,Athanassopoulos:1996jb,%
Athanassopoulos:1998pv,Athanassopoulos:1998er,Aguilar:2001ty,Declais:1995su,Apollonio:1999ae,Boehm:2000gk,%
Eitel:2000by,Armbruster:2002mp,Avvakumov:2002jj}.
However, neutrino oscillations are sensitive only to the mass (squared) splittings 
$\triangle m_{ij}^2 = m_{\nu_i}^2 - m_{\nu_j}^2$, not to the individual masses, 
$(m_{\nu_4}>)\ m_{\nu_3}>m_{\nu_2}>m_{\nu_1}$, themselves.
Only a lower bound on the mass of the heaviest neutrino can be derived from these observations, e.g. 
\begin{equation}
\label{lim_low_atm}
m_{\nu_3}\geq\sqrt{\triangle m_{\rm atm}^2}\,\gwig\, 0.04\ {\rm eV}
\end{equation}
from the atmospheric mass splitting in a three neutrino flavour scenario and
\begin{equation}
\label{lim_low_lsnd}
m_{\nu_4}\geq\sqrt{\triangle m_{\rm LSND}^2}\,\gwig\, 0.4\ {\rm eV}
\end{equation}
from the still allowed value of the LSND mass splitting in a four flavour scenario, respectively (for a 
recent compilation of global analyses on mass splitting and mixing parameters, see for example 
Refs.~\cite{Fogli:2001vr,Bahcall:2001zu,Gonzalez-Garcia:2002dz} and references therein).

For an investigation of 
the absolute scale of neutrino masses one has to exploit different types of experiments, 
such as the search for mass imprints in the endpoint spectrum of tritium beta ($\beta$) decay or the search for 
neutrinoless double beta ($0\nu 2\beta$) decay.  
At present, these direct kinematical measurements of neutrino masses provide only upper limits, e.g. 
\begin{equation}
\label{lim_beta}
m_\beta\equiv \sqrt{\sum_i |U_{ei}|^2\, m_{\nu_i}^2 }<2.2\div 2.5\ {\rm eV}
\hspace{2ex} (95\,\%\ {\rm C.L.})
\,,
\end{equation} 
with $U$ being the leptonic mixing matrix,  
for the effective mass measured in tritium $\beta$ 
decay~\cite{Weinheimer:1999tn,Lobashev:1999tp,Bonn:2001tw,Lobashev:2001uu}, 
and 
\begin{equation}
\langle m_\nu\rangle \equiv \left| \sum_i U_{ei}^2\, m_{\nu_i} \right| <0.33\div 1.35\ {\rm eV}
\hspace{2ex} (90\,\%\ {\rm C.L.})
\end{equation}
for the Majorana neutrino mass 
parameter~\cite{Aalseth:1999ji,Baudis:1999xd,Aalseth:2000ud,Klapdor-Kleingrothaus:2001yx,Aalseth:2002rf}
appearing in $0\nu 2\beta$ 
decay. The recently reported evidence for 
neutrinoless double beta decay and correspondingly deduced parameter range~\cite{Klapdor-Kleingrothaus:2002ke} 
\begin{equation}
\label{m0n2b}
\langle m_\nu\rangle =(0.11\div 0.56)\ {\rm eV}
\hspace{2ex} (95\,\%\ {\rm C.L.})
\end{equation}
has been seriously challenged by Refs.~\cite{Feruglio:2002af,Aalseth:2002dt}. 
Combining the experimental constraints from oscillations and from tritium $\beta$ decay, one infers
upper bounds on the mass of the heaviest neutrino~\cite{Barger:1998kz},
\begin{equation}
\label{lim_comb_osc_beta}
m_{\nu_3}<\sqrt{m_\beta^2+\triangle m_{\rm atm}^2}\,\lwig\, 2.5\ {\rm eV}\,, 
\end{equation}
in a three flavour, and 
\begin{equation}
\label{lim_comb_osc_beta_lsnd}
m_{\nu_4}<\sqrt{m_\beta^2+\triangle m_{\rm LSND}^2}\,\lwig\, 3.8\ {\rm eV}\,, 
\end{equation}
in a four flavour scenario, respectively. For further recent investigations of the relationship 
between oscillation phenomena, $\beta$ decay and 
$0\nu 2\beta$ decay, in particular with respect to the neutrino mass spectrum, one may also consult 
Refs.~\cite{Klapdor-Kleingrothaus:2001gr,Vissani:2001ci,Bilenky:2001rz,Bilenky:2001xq,Farzan:2001cj,%
Osland:2001pm,Czakon:2002uh,Pascoli:2001vr} 
and references cited 
therein.

Further information on the absolute scale of neutrino masses can be obtained through cosmological and astrophysical considerations.
Neutrinos in the $0.1\div 1$~eV mass range have cosmological implications 
since they represent a non-negligible part of dark matter. This gives the opportunity to put upper limits
on neutrino masses from cosmology~\cite{Gershtein:1966gg,Dolgov:2002wy}. Analyses of galaxy clustering,
including recent CMB measurements and other cosmological constraints,  
give an upper bound 
\begin{equation}
\label{lim_cosm}
\sum_i 
m_{\nu_i} < 1.8\div 4.4\ {\rm eV}
\end{equation}
on the sum of the neutrino 
masses~\cite{Croft:1999mm,Fukugita:2000as,Gawiser:2000yg,Wang:2001}.
From the spread of arrival times of neutrinos from supernova SN 1987A, coupled with the measured 
neutrino energies,  a time-of-flight limit of  
$m_\beta <23$~eV can be derived~\cite{Loredo:1988mk,Kernan:1995kt}, which, however, is not competitive 
with the direct limit~(\ref{lim_beta}).  

It is extremely welcome that the Z-burst scenario opens a further and timely window to the 
absolute neutrino mass scale, since the opportunities to determine this crucial quantity are rare at present. 
In addition, its verification would give us an indirect detection of the so elusive relic neutrinos 
and, finally, offer an explanation of the origin of the highest energy cosmic rays. 

The organization of the present paper is as follows. In Section~\ref{sect:spectra} we 
describe our determination of the proton and photon spectra at Earth, 
which originate 
from Z-bursts taking place mainly at extragalactic distances. We discuss in detail the 
main ingredients in the predictions of the spectra, such as the details of Z production and hadronic decay, 
the propagation of nucleons and photons through the diffuse extragalactic photon background, the 
diffuse flux of ultrahigh energy cosmic neutrinos, and the relic neutrino number density, as well as their 
anticipated uncertainties. The comparison of the Z-burst spectra with the observed ultrahigh energy
cosmic ray spectrum is 
presented in Section~\ref{sect:determination}. The required absolute neutrino mass, as well as the 
necessary ultrahigh energy cosmic neutrino flux, are determined via a maximum likelihood analysis for various 
assumptions about the nature of the background of ordinary cosmic rays  
from unresolved astrophysical sources and variations of
the diffuse extragalactic photon background notably in the radio band.
On the basis of these studies we find that the Z-burst determinations of the mass of the heaviest neutrino 
as well as of the neutrino flux are fairly robust. We find a required neutrino mass range of 
\massrangeegonesig\ 
at the 68\,\% confidence level, if the background of ordinary cosmic rays is of extragalactic origin.  
This range narrows down considerably if a particular universal radio background is assumed, e.g. 
to \massrangeegstrongonesig\ for a large one. 
In Section~\ref{sect:discussion} we discuss the implications of our findings for future laboratory studies
such as the tritium beta decay and  
neutrinoless double beta decay, as well as for astrophysical and cosmological neutrino investigations.
This section contains also our conclusions.     

\section{\label{sect:spectra}Z-burst spectra}

Our comparison of the Z-burst scenario with the observed UHE\-CR spectrum
proceeds as follows. First, we determine the probability
of Z production as a function of the distance from Earth.
Secondly, we exploit collider experiments to derive 
the energy distribution of the produced protons and photons in the laboratory (lab) system.
Thirdly, we consider the propagation of the protons and photons, i.\,e. we determine 
their energy losses due to pion and/or $e^+e^-$ production 
through scattering on the diffuse extragalactic background photons and due to their redshift. 
The last step is the comparison of the predicted and the observed spectra and 
the extraction of the required mass of the heaviest relic neutrino and of the necessary UHEC$\nu$ flux.    

Our prediction of the differential proton flux, i.e. the number of protons 
arriving at Earth with energy $E$ per units of energy ($E$), of area ($A$), of time ($t$) and
of solid angle ($\Omega$),  
\begin{equation}
F_{p|Z} ( E ) = \frac{{\rm d}N_{p|Z}}{{\rm d}E\,{\rm d}A\,{\rm d}t\,{\rm d}\Omega}
\,,
\end{equation} 
from Z-bursts  can be summarized as
\begin{eqnarray}
\label{p-flux}
\lefteqn{ F_{p|Z} ( E ) =  \sum_{i}
\int\limits_0^\infty {\rm d}E_p \int\limits_0^{R_{\rm max}} {\rm d}r 
 }  
\\[1ex] \nonumber && \times
\left[  \int\limits_0^\infty {\rm d} E_{\nu_i}\,F_{\nu_i}(E_{\nu_i},r)\, n_{\bar\nu_i}(r)
\right.
\\[1ex]\nonumber && \left.
\hspace{6ex}
+ \int\limits_0^\infty {\rm d} E_{\bar\nu_i}\, F_{\bar\nu_i}(E_{\bar\nu_i},r)\, n_{\nu_i}(r)
\right]
\\[1ex] \nonumber && \times
 \sigma_{\nu_i\bar\nu_i}(s)\,
{\rm Br}(Z\to {\rm hadrons})\,
\frac{{\rm d} N_{p+n}}{{\rm d} E_p}
\\[1ex] \nonumber && \times
(-)\frac{\partial}{\partial E} P_p(r,E_p;E)
\,,
\end{eqnarray}
with the following important building blocks:
the UHEC$\nu$ fluxes $F_{\nu_i}(E_{\nu_i},r)$ at the energies 
$E_{\nu_i}$ ($\approx E_{\nu_i}^{\rm res}$) and at the distance $r$
of Z production to Earth,  the number density 
$n_{\nu_i}(r)$ of the 
relic neutrinos, 
the Z production cross section $\sigma_{\nu_i\bar\nu_i}(s)$ at centre-of-mass (cm)
energy squared $s = 2\,m_{\nu_i}\,E_{\nu_i}$, the branching ratio 
${\rm Br}(Z\to {\rm hadrons})$, 
the energy distribution ${\rm d} N_{p+n}/{\rm d} E_p$ of the produced protons 
(and neutrons) with energy $E_p$, and the probability $P_p(r,E_p;E)$ 
that a proton created at a distance $r$ with energy $E_p$ arrives
at Earth above the threshold energy $E$. 

A similar expression as Eq.~(\ref{p-flux}) holds for the differential
photon flux from Z-bursts, 
\begin{eqnarray}
\label{ph-flux}
\lefteqn{ F_{\gamma |Z} ( E ) =  \sum_{i}
\int\limits_0^\infty {\rm d}E_\gamma \int\limits_0^{R_{\rm max}} {\rm d}r 
 }  
\\[1ex] \nonumber && \times
\left[  \int\limits_0^\infty {\rm d} E_{\nu_i}\,F_{\nu_i}(E_{\nu_i},r)\, n_{\bar\nu_i}(r)
\right.
\\[1ex] \nonumber && \left. 
\hspace{6ex}
+  \int\limits_0^\infty {\rm d} E_{\bar\nu_i}\,F_{\bar\nu_i}(E_{\bar\nu_i},r)\, n_{\nu_i}(r)
\right]
\\[1ex] \nonumber && \times
 \sigma_{\nu_i\bar\nu_i}( s)\,
{\rm Br}(Z\to {\rm hadrons})\,
\frac{{\rm d} N_\gamma}{{\rm d} E_\gamma}
\\[1ex] \nonumber && \times
(-)\frac{\partial}{\partial E} P_\gamma (r,E_\gamma ;E)
\,.
\end{eqnarray} 
Here, the photon propagation function $P_\gamma (r,E_\gamma ;E)$ gives the
expected number of detected photons above the threshold 
energy $E$ if one photon started from a distance of $r$ with energy $E_\gamma$.
Note, that the $P_\gamma$ function has a different 
interpretation than the $P$-function of protons. This arises from the fact that the number 
of photons -- in distinction to the number of protons -- is not conserved during their propagation.  

The building blocks related to Z-production and decay -- $\sigma_{\nu_i\bar\nu_i}$, the hadronic branching ratio, and 
the momentum distributions ${\rm d} N_i/{\rm d} E_i$ -- are very well, and the propagation functions $P_i$ are fairly well determined, 
whereas the first two ingredients, the flux of UHEC$\nu$s, $F_{\nu_i}(E_{\nu_i},r)$, and the radial distribution of
the relic neutrino number density $n_{\nu_i}(r)$, are much less accurately known. 
In the following we shall discuss all these ingredients in detail.

\subsection{\label{zproddecay}Z production and decay}

At LEP and SLC millions of Z bosons 
were produced and their decays analyzed with extreme high accuracy. 
Due to the large statistics, the uncertainties of our analysis
related to Z decay turned out to be negligible.  

Let us start with a review of the Standard Model neutrino annihilation cross section.
The s-channel Z-exchange annihilation cross section into any fermion anti-fermion ($f\bar f$) pair 
is given by (see e.\,g. Refs.~\cite{Roulet:1993pz,Gondolo:1993rn})
\begin{equation}
\label{z-res-cs-roulet}
\sigma_{\nu_i \bar{\nu}_i}^{Z} (s)
=
\frac{G_F^2\,s}{4\,\pi}\, \frac{M_Z^4}{(s-M_Z^2)^2+M_Z^2\, \Gamma_Z^2} 
\ N_{\rm eff}(s)\,,
\end{equation}
where $s$ is the cm energy squared, $\Gamma_Z$ is the total width of the Z boson, 
and $N_{\rm eff}$ is the effective number of annihilation channels,
\begin{eqnarray}
\lefteqn{ 
N_{\rm eff}(s) = 
\sum_f \theta (s-4\,m_f^2)\,\times
}
\\[1ex] \nonumber
&& \times \, \frac{2}{3}\,n_f \left( 1 - 
8\, T_{3\,f}\, q_f\, \sin^2\theta_W
+8\,q_f^2\sin^4\theta_W\right)
\,.
\end{eqnarray}
Here the sum is over all fermions with $m_f<\sqrt{s}/2$, with charge
$q_f$ (in units of the proton charge), isospin $T_{3\,f}$ ($1/2$ for $u,c$ and 
neutrinos; $-1/2$ for $d,s,b$ and negatively charged leptons), and $n_f=1(3)$ 
for leptons (quarks ($q$)). With $\sin^2\theta_W=0.23147(16)$ for the $\sin^2$ of
the effective Weinberg angle and $G_F=1.16639(1)\times 10^{-5}$ GeV$^{-2}$ 
for the Fermi coupling constant~\cite{Groom:2000in}, 
formula~(\ref{z-res-cs-roulet}) gives, at the $Z$-mass,
\begin{eqnarray}
\sigma (\nu_i \bar{\nu}_i \to Z^\ast \to
 {\rm all\ }q\bar q)\mid_{s=M_Z^2} &=&
314.9\ {\rm nb}\,,
\\[1ex] 
\label{cs-tot-zpeak}
\sigma (\nu_i \bar{\nu}_i \to Z^\ast \to
 {\rm all}\ f\bar f)\mid_{s=M_Z^2} &=&
455.6\ {\rm nb}\,,
\end{eqnarray}
with a branching ratio $\sigma (\nu_i \bar{\nu}_i \to Z^\ast
\to {\rm all\ }q\bar q)/\sigma (\nu_i \bar{\nu}_i \to Z^\ast
\to {\rm all}\ f\bar f)\mid_{s=M_Z^2} =  0.6912$, in good agreement
with the experimental result~\cite{Groom:2000in}, 
\begin{equation}
{\rm Br}(Z\to {\rm hadrons})=(69.89\pm 0.07)\,\%\,.
\end{equation}  

Later we shall exploit the following simplification which arises due to the fact that 
the cross section~(\ref{z-res-cs-roulet}) is sharply peaked at the resonance
cm energy squared $s=M_Z^2$. Correspondingly, it acts essentially like a $\delta$-function in 
the integration over the energies $E_{\nu_i}$ of the incident neutrinos in Eqs.~(\ref{p-flux}) and 
(\ref{ph-flux}), and we can assume that the UHEC$\nu$ fluxes are constant in the relevant energy
region. Thus, introducing the energy-averaged annihilation 
cross section~\cite{Weiler:1982qy,Weiler:1999sh},
\begin{equation}
\label{sig_ann}
\langle \sigma_{\rm ann}\rangle 
\equiv
\int
\frac{{\rm d}s}{M_Z^2}\,\sigma_{\rm ann} (s)
= 2\,\pi\,\sqrt{2}\,G_F
= 40.4\ {\rm nb}\,,  
\end{equation}
which is the effective cross section for all neutrinos within 
$1/2\,\delta E_{\nu_i}^{\rm res}/E_{\nu_i}^{\rm res}=\Gamma_Z/M_Z=2.7\,\%$ of their
peak annihilation energy, we can write
\begin{eqnarray}
\label{eresfres}
\lefteqn{
\int\limits_0^\infty {\rm d} E_{\nu_i}\,F_{\nu_i}(E_{\nu_i})\,\sigma_{\nu_i\bar\nu_i}( s=2\,m_{\nu_i}\,E_{\nu_i})
        }
\\[1ex]\nonumber && \hspace{8ex}
\simeq 
E_{\nu_i}^{\rm res}\,F_{\nu_i}(E_{\nu_i}^{\rm res})\,\langle \sigma_{\rm ann}\rangle 
\,.
\end{eqnarray}

In view of the expected rapid decrease of the UHEC$\nu$ flux at increasing energies 
(cf. Section~\ref{fluxes}),
we neglect $t$-channel W- and Z-exchange annihilation processes. 
On resonance, the $s$-channel Z-exchange processes completely overwhelm them. 
Fairly above the resonant energies, on the other hand, they eventually dominate
but are probably unobservable due to the lack of an appreciable UHEC$\nu$ flux.  

\begin{figure}
\begin{center}
\includegraphics[bbllx=20pt,bblly=225pt,bburx=570pt,bbury=608pt,width=8.65cm]
{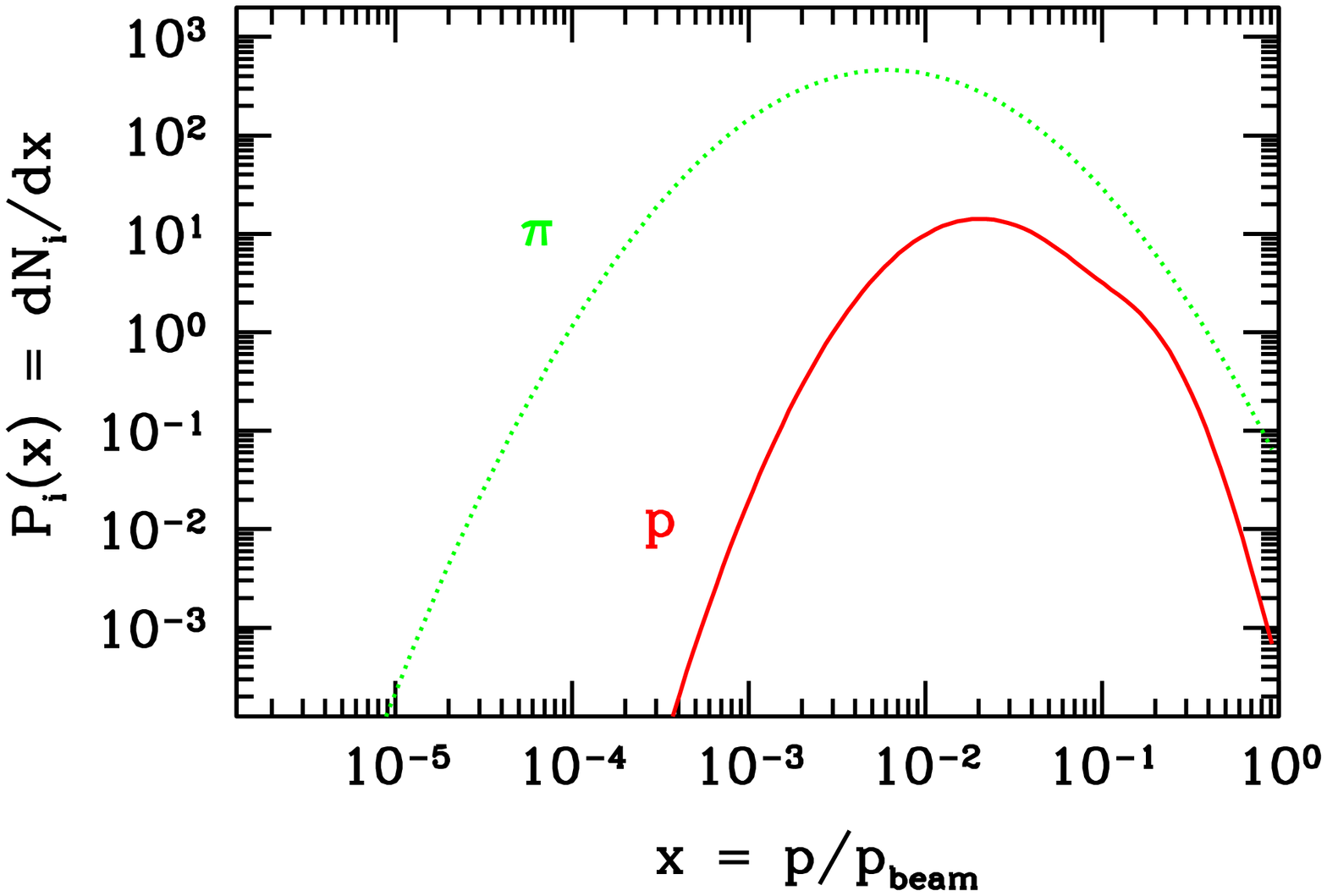}
\includegraphics[bbllx=20pt,bblly=225pt,bburx=570pt,bbury=608pt,width=8.65cm]
{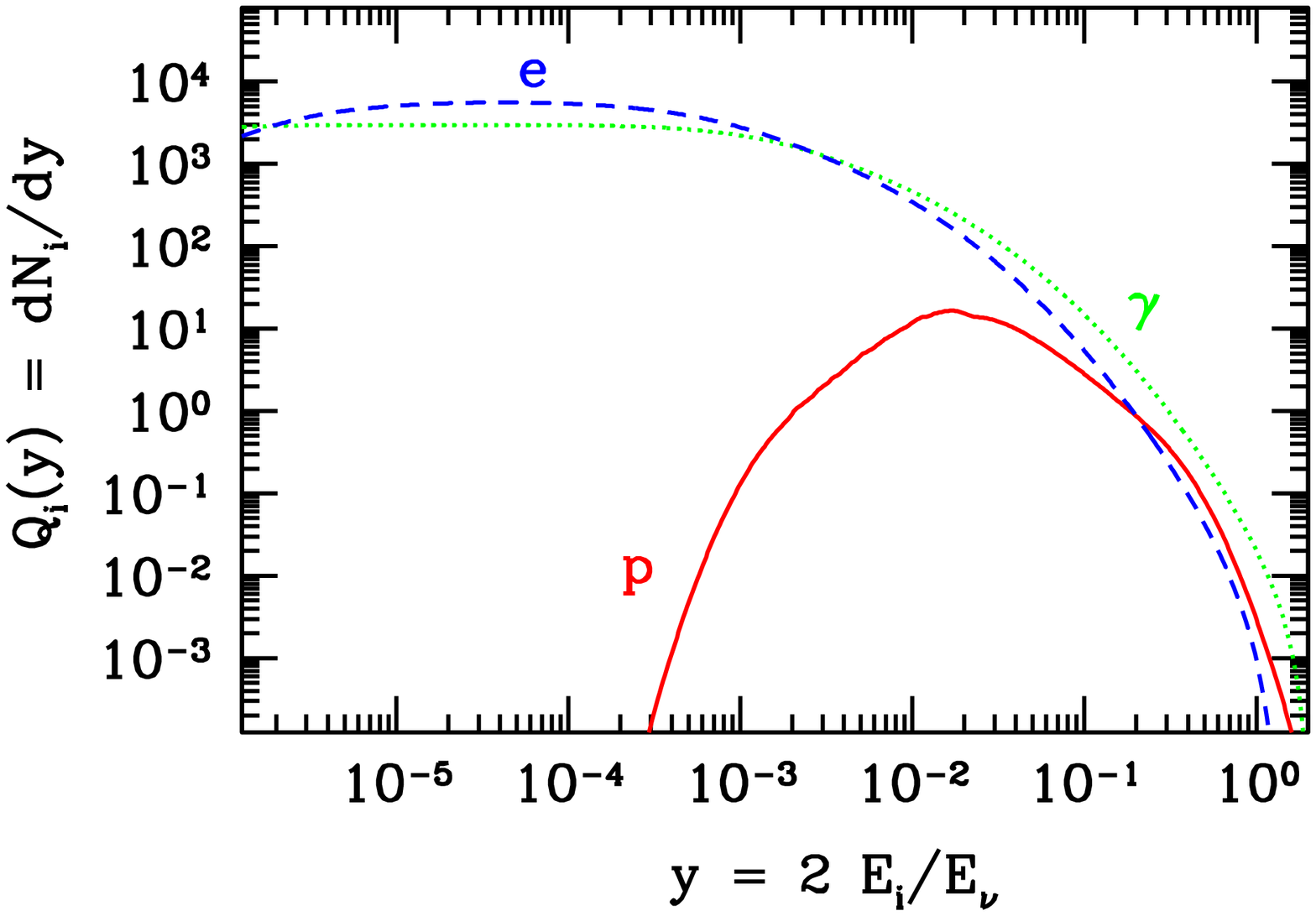}
\caption[...]{\label{prot-mom-dist}
Momentum distributions in hadronic Z decays. 
{\em Top:} Combined data from Refs.~\cite{Akers:1994ez,Abreu:1995cu,Buskulic:1994ft,Abe:1999qh,Opal:unp}
on proton (plus antiproton) momentum distribution (solid), 
normalized to $\langle N_p\rangle =1.04$, and charged pion momentum distributions (dotted), 
normalized to $\langle N_{\pi^\pm}\rangle = 16.99$.
{\em Bottom:} Distribution of protons (``p''; solid), photons (``$\gamma$''; dotted) and 
electrons (``e''; dashed) in the lab system, in which the target 
relic neutrino is at rest. 
}
\end{center}
\end{figure}

Next, we turn to the energy distribution of the protons and photons 
in Z decay. We combined 
existing published and some improved unpublished data on the momentum 
distribution, 
\begin{equation}
{\cal P}_p\,(x)\equiv \frac{{\rm d}N_p}{{\rm d}x}\,,
\hspace{6ex}
x\equiv \frac{p_p}{p_{\rm beam}}\,,
\end{equation}
of protons ($p$) (plus antiprotons ($\bar p$)) in Z 
decays~\cite{Akers:1994ez,Abreu:1995cu,Buskulic:1994ft,Abe:1999qh,Opal:unp}, see Fig.~\ref{prot-mom-dist} (top). 
The experimental data, ranging down to $x\approx 8\cdot 10^{-3}$, were 
combined with the predictions from the modified leading logarithmic approximation (MLLA)~\cite{Azimov:1985np,Azimov:1986by,Fong:1991nt}
at low $x$.
The $p+{\bar p}$ multiplicity is 
$\langle N_p\rangle = \int_0^1 {\rm d}x\, {\cal P}_p(x) = 1.04\pm 0.04$ in the 
hadronic channel~\cite{Groom:2000in}. 

In the cm system of the Z production the
angular distribution of the hadrons is determined by the
spin $1/2$ of the primary quarks and thus proportional to 
$1+w^2=1+\cos^2\theta$ (here $\theta$ is the angle between the
incoming neutrinos and the outgoing hadrons (cf.~\cite{Schmelling:1995py})).  
The energy distribution of the produced protons with energy 
$E_p$ entering the Z-burst spectrum~(\ref{p-flux}),
\begin{equation}
\label{Q-def}
\frac{{\rm d} N_{p}}{{\rm d} E_p} = 
\frac{2}{E_\nu}\,\frac{{\rm d} N_p}{{\rm d} y}
\equiv
\frac{2}{E_\nu}\,{\cal Q}_p(y)\,,
\end{equation}
with $y=2E_p/E_\nu$,
is finally obtained after a Lorentz transformation from the cm system
to the lab system,
\begin{eqnarray}
\label{Q-dist}
&&{\mathcal Q}_p(y)= \sum_{+,-} {3 \over 8} \int_{-1}^{+1} {\rm d}w\, (1+w^2)  \\
&&\times\, {1 \over 1-w^2}
\left|
{{ \pm y-w\sqrt{y^2-(1-w^2)(2m_p/M_Z)^2}}
\over \sqrt{y^2-(1-w^2)(2m_p/M_Z)^2}}
\right|
\nonumber\\
&&{\cal P}_p\left( \frac{-wy \pm \sqrt{y^2-(1-w^2)(2m_p/M_Z)^2}}{1-w^2} \right), 
\nonumber
\end{eqnarray}
where $m_p$ is the proton mass.  
The first line comes from the angular distribution, 
the second line is the Jacobian and
the third one is the momentum distribution at the inverted momentum.
The scaled energy distribution ${\cal Q}_p$, as defined in Eq.~(\ref{Q-def}) and given by Eq.~(\ref{Q-dist}), is 
displayed in Fig.~\ref{prot-mom-dist} (bottom). 

Neutrons produced in Z decays will decay and end up as UHECR protons. 
They are taken into account according to 
\begin{equation}
{\mathcal Q}_{p+n}(y)=
\left(1+\frac{\langle N_n\rangle}{\langle N_p\rangle}\right)\,{\cal Q}_p(y)\,,
\end{equation}
where the neutron ($n$) (plus antineutron ($\bar n$)) multiplicity, 
$\langle N_n\rangle = \int_0^2 {\rm d}y\, {\cal Q}_n(y)$, is $\approx 4\%$ 
smaller than the proton's~\cite{Biebel01}. 

Photons are produced in hadronic Z decays via fragmentation into neutral pions, 
$Z\to \pi^0 + X\to 2\,\gamma + X$ (cf. Fig.~\ref{illu}). 
The corresponding scaled energy distribution in the lab system, defined analogously to 
Eq.~(\ref{Q-def}), reads
\begin{equation}
\label{Q-dist-gamma}
{\mathcal Q}_\gamma (y)
=\int_{-1}^{1} {\rm d}w\,  
\frac{2}{1+w}\,
{\mathcal Q}_{\pi^0} 
\left( \frac{2\,y}{1+w} \right) 
\,,
\end{equation}
where the scaled energy distribution ${\mathcal Q}_{\pi^0}(y)$, with $y=2\,E_{\pi^0}/E_\nu$, 
is given by Eq.~(\ref{Q-dist}), with $m_p\to m_{\pi^0}$ and  ${\mathcal P}_p\to {\mathcal P}_{\pi^0}$, 
the momentum distribution of pions in hadronic Z decay. For the latter distribution, we take the measured one
of charged pions ${\mathcal P}_{\pi^\pm}$ from hadronic Z decay~\cite{Akers:1994ez,Abreu:1995cu,Buskulic:1994ft,Abe:1999qh,Opal:unp}
(cf. Fig.~\ref{prot-mom-dist} (top)), 
normalized such that $\langle N_\gamma\rangle = \int_0^2 {\rm d}y\, {\cal Q}_\gamma (y)= 20.97$~\cite{Groom:2000in}.  
Note, that the distributions ${\mathcal Q}_i$, $i=p,\gamma$, presented in Fig.~\ref{prot-mom-dist} (bottom), 
compare favorably with the ones presented in Ref.~\cite{Gelmini:2002xy} based on the event generator 
PYTHIA~\cite{Sjostrand:1994yb}. 

Electrons (and positrons) from hadronic Z decay are also relevant for the development of electromagnetic cascades.
They stem from decays of secondary charged pions, $Z\to \pi^\pm +X\to e^\pm + X$ (cf. Fig.~\ref{illu}), 
and their scaled energy distribution in terms of $y=2\,E_e/E_\nu$ reads 
\begin{eqnarray}
{\mathcal Q}_{e^\pm}(y) &=&
\int_{-1}^{1}dw \int_0^2 dx 
\frac{1}{xw+\sqrt{x^2+(2\,m_e/m_\pi)^2}}
\\[1ex] \nonumber && \times
{\mathcal Q}_{\pi^\pm}
\left( 
\frac{2\, y}{xw+\sqrt{x^2+(2\,m_e/m_\pi)^2}} \right) 
{\mathcal P}_e (x)
\,,
\end{eqnarray}
where ${\mathcal P}_e$ is the momentum distribution of the electrons in the rest system of the 
charged pion. The energy distribution ${\mathcal Q}_{e^\pm}$ is also 
displayed in Fig.~\ref{prot-mom-dist} (bottom).

\subsection{\label{prop}Propagation of nucleons and photons 
through the diffuse extragalactic photon background}

The cosmic microwave background 
is known to a high accuracy.  It plays the key role in the 
determination of the probability $P_p(r,E_p;E)$
that a proton created at a distance $r$ with energy $E_p$ arrives
at Earth above the threshold energy $E$, suggested in Ref.~\cite{Bahcall:2000ap} and
determined for a wide range of parameters in Ref.~\cite{Fodor:2001yi}. 
The propagation function $P_p$ takes into 
account the fact that protons of extragalactic (EG) origin and energies above 
$\approx 4\cdot 10^{19}$ eV 
lose a large fraction of their energy
due to pion and $e^+e^-$ production 
through scattering on the CMB and due to their redshift~\cite{Greisen:1966jv,Zatsepin:1966jv}.
In the present analysis we have included new results for $P_p(r,E_p;E)$ which 
include now variations in the cosmological parameters,  
in extension to the already published form~\cite{Fodor:2001yi} 
exploited in Ref.~\cite{Fodor:2001qy}. 
In our analysis we go, according to 
\begin{equation}
\label{z-H}
{\rm d}z = - (1+z)\,H(z)\,{\rm d}r/c\,,
\end{equation}
out to distances $R_{\rm max}$ (cf.~(\ref{p-flux})) 
corresponding to redshift $z_{\rm max} = 2$ (cf.~Ref.~\cite{Waxman:1995dg}). 
We use the expression 
\begin{equation}
\label{H-Omega}
H^2(z) = H_0^2\,\left[ \Omega_{M}\,(1+z)^3 
+ \Omega_{\Lambda}\right]
\end{equation} 
for the relation of the Hubble expansion rate at redshift $z$ to the present one.
Uncertainties of the latter, $H_0=h$ 100 km/s/Mpc, with 
$h=(0.71\pm 0.07)\times^{1.15}_{0.95}$~\cite{Groom:2000in}, 
are included. 
In Eq.~(\ref{H-Omega}), $\Omega_{M}$ and $\Omega_{\Lambda}$, with $\Omega_M+\Omega_\Lambda =1$, are the present 
matter and vacuum energy densities in terms of the critical density. As default values we choose
$\Omega_M = 0.3$ and $\Omega_\Lambda = 0.7$, as favored today. Our results
turn out to be pretty insensitive to the precise values of the cosmological parameters.

\begin{figure}
\begin{center}
\includegraphics[bbllx=20pt,bblly=221pt,bburx=570pt,bbury=608pt,width=8.65cm]
{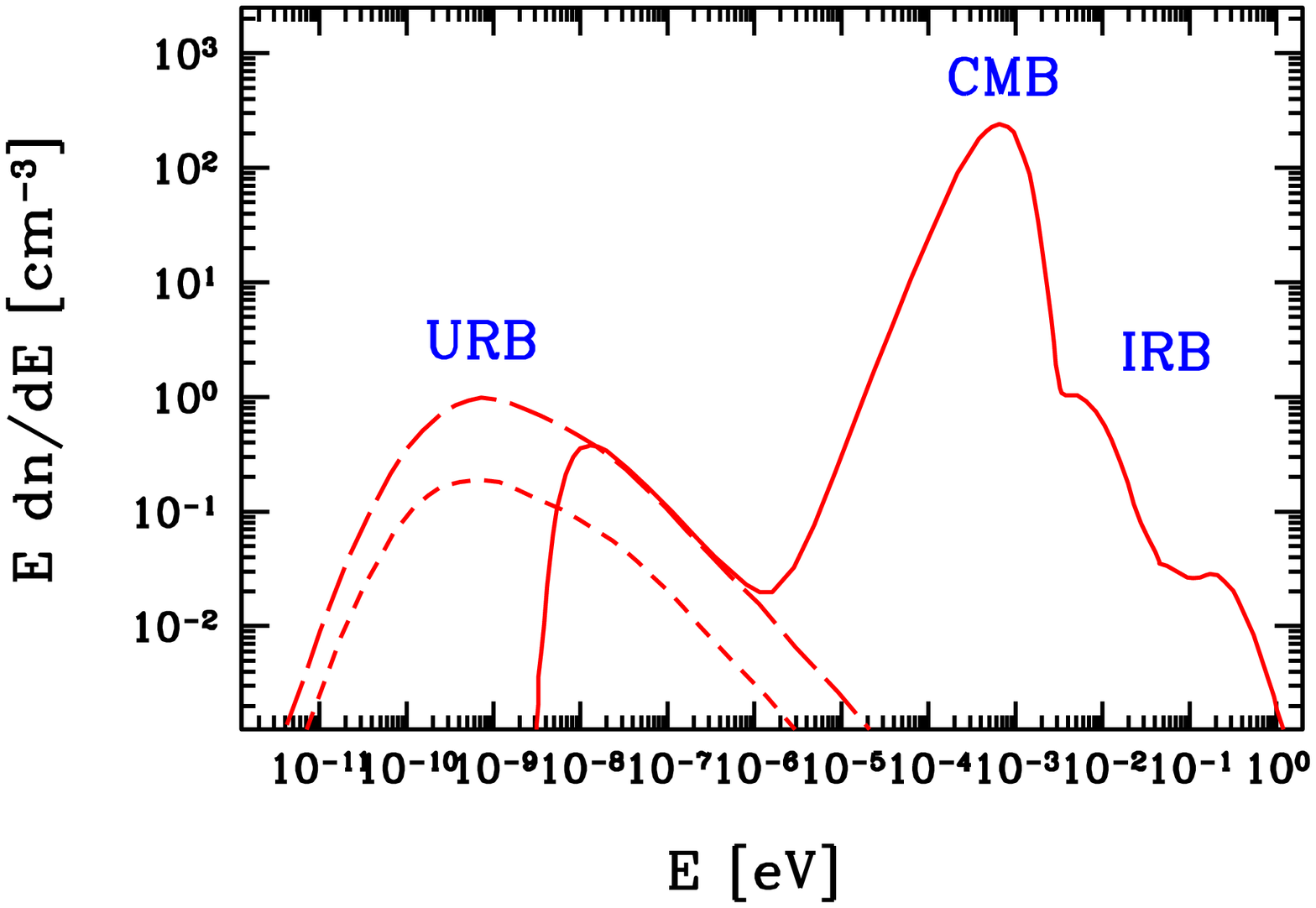}
\includegraphics[bbllx=20pt,bblly=221pt,bburx=570pt,bbury=608pt,width=8.65cm]
{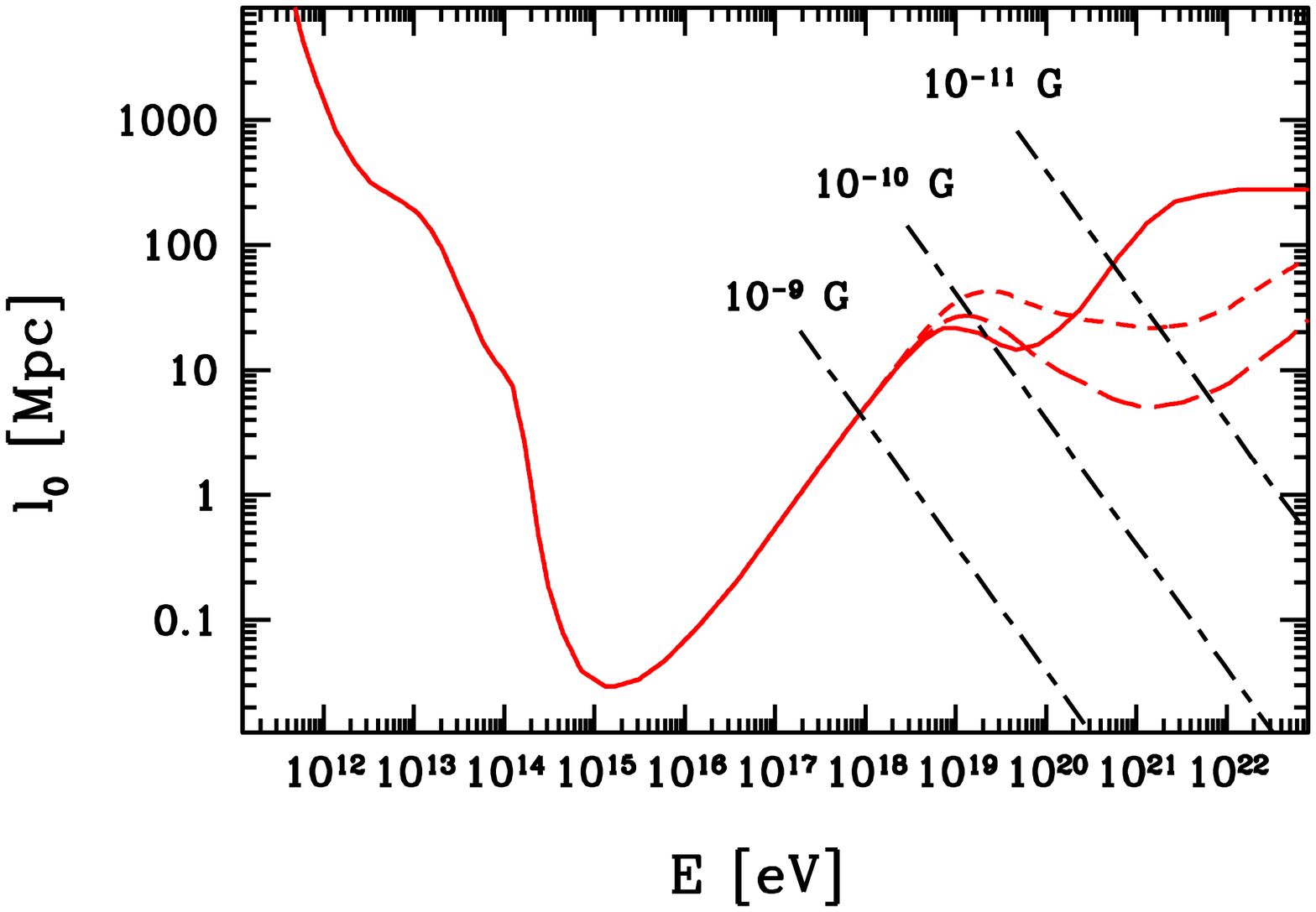}
\caption[...]{\label{ph_mfp}
{\em Top:} The intensity spectrum of the diffuse extragalactic photon background 
at redshift $z=0$ from Ref.~\cite{Lee:1998fp} (solid). 
Different estimates from Ref.~\cite{Protheroe:1996si} of the universal radio 
background (URB) are also indicated: 
high URB (long-dashed) and moderate URB (short-dashed). 
{\em Bottom:} Photon energy attenuation length $l_0$ at $z=0$ corresponding to the photon 
background shown above ~\cite{Lee:1998fp} (solid).
Variations of $l_0$ arising from different assumptions about the URB (cf. Ref.~\cite{Protheroe:1996si} and
top) are also indicated: high URB (long-dashed) and moderate URB (short-dashed).
Shown is furthermore the energy attenuation 
length for electrons due to synchrotron radiation (long-dashed-short-dashed), for different 
magnitudes of the extragalactic magnetic fields.
}
\end{center}
\end{figure}

The simulations needed for the computation of $P_p(r,E_p;E)$ require large
computer power. We have used a farm of personal computers 
(PCs) consisting of 128 parallel $1.7$~GHz 
Pentium 4 processors -- normally devoted to QCD lattice calculations~\cite{Fodor:2002zi} -- 
and have exploited about 3~hours CPU time to determine $P_p(r,E_p;E)$ in a range
of $r\leq 4000$~Mpc, $10^{18}$~eV $\leq E_p\leq 10^{26}$~eV, and $10^{18}$~eV $\leq E\leq 10^{26}$~eV, 
for each  fixed value of the cosmological parameters.
The simulation was carried out in small ($10$~kpc) steps in $r$. For each step, 
the statistical energy losses due to pion/$e^+e^-$ production and redshift are taken 
into account~\cite{Fodor:2001yi}. 
In this connection, the advantage of our formulation of the Z-burst spectrum (cf.~(\ref{p-flux}))
in terms of the probability  $P_p(r,E_p;E)$ becomes evident. We have to determine the latter       
only once and for all. Without the use of $P_p(r,E_p;E)$, we would have to perform a simulation
for any variation of the input spectrum, notably for any change in the neutrino mass. 
Our maximum likelihood analysis, involving the neutrino mass as a free parameter, would
thus require excessive computer power -- on the order of 300 hours on the above PC farm for fixed
cosmological parameters.   
Since $P_p(r,E_p;E)$ is of universal usage, we have decided to make the corresponding  
numerical data for the probability distribution 
$(-)\partial P_p (r,E_p;E)/\partial E$   
available for the public via the World-Wide-Web URL 
\begin{center}
{\it http://www.desy.de/\~{}uhecr}  \,.  
\end{center}

The determination of the photon propagation function $P_\gamma (r,E_\gamma ;E)$,
entering the photon flux prediction~(\ref{ph-flux}), was done as follows. 
In distinction to the case of the proton propagation function, we used here 
the continuous energy loss  (CEL) approximation which 
largely simplifies the work and reduces the required computer resources -- we have estimated
the necessary CPU time for a full simulation on the PC farm mentioned above to 1000~hours per
fixed set of the cosmological parameters.  
In the CEL 
approximation, the energy (and number) of the detected photons is a unique function of
the initial energy and distance, and statistical fluctuations are neglected.
A full simulation of the photon propagation function will be the subject of a later
work.   

The processes that are taken into account are pair production on the diffuse extragalactic 
photon background (cf. Fig.~\ref{ph_mfp} (top)), double pair production and inverse Compton scattering of
the produced pairs. We comment also on synchrotron radiation in a possible extragalactic magnetic field (EGMF).  
For the energy attenuation length of the photons due to these processes, we exploited the values quoted in  
Ref.~\cite{Lee:1998fp} (see also Ref.~\cite{Protheroe:1996ft}) and the further ones presented in
Fig.~\ref{ph_mfp} (bottom) which incorporate various assumptions
about the poorely known universal radio (URB) (from Ref.~\cite{Protheroe:1996si}) and infrared (IRB) backgrounds. 
We shall analyse later the dependence of the neutrino mass and other fit parameters 
on these variations. Note, that, in view of the recent URB estimates in 
Ref.~\cite{Protheroe:1996si}, the ones presented in Fig.~\ref{ph_mfp} (top)), 
which are based on Ref.~\cite{Clark:1970}, can be referred to as ``minimal'' URB.  

The computation of the photon propagation function $P_\gamma (r,E_\gamma ;E)$ was carried out in the following way.
The energy attenuation of photons in the CEL approximation 
was calculated according to 
\begin{eqnarray}
\label{E-atten}
dE=-E\left(\frac{{\rm d}r}{l_z(E)}-\frac{{\rm d}z}{1+z}\right)\,,
\end{eqnarray}
where $l_z(E)=(1+z)^{-3}\, l_0(E(1+z))$ is the energy attenuation length at redshift $z$.
The number of photons was assumed to be constant at ultrahigh energies $\gwig 10^{18}$~eV, 
due to the small inelasticities in this energy range. Below, it was increased in a way to 
maintain energy conservation (except for the redshift contribution):
\begin{eqnarray}
\label{n-phot}
{\rm d}N_\gamma =-N_\gamma\, \frac{{\rm d}r}{l_z(E)}\,.
\end{eqnarray}
The $P_\gamma (r,E_\gamma ;E)$ function was then obtained by integration of these equations.
In the ultrahigh energy region -- which is most relevant for us since we perform our fit to the 
cosmic ray data there --  
the approximation described above gives the photon flux quite reliable, while at lower energies 
it yields an upper bound.

\subsection{\label{fluxes}UHEC\boldmath$\nu$\unboldmath\ fluxes}

\begin{figure}
\vspace*{2.0mm} 
\includegraphics[bbllx=20pt,bblly=221pt,bburx=570pt,bbury=608pt,width=8.0cm]
{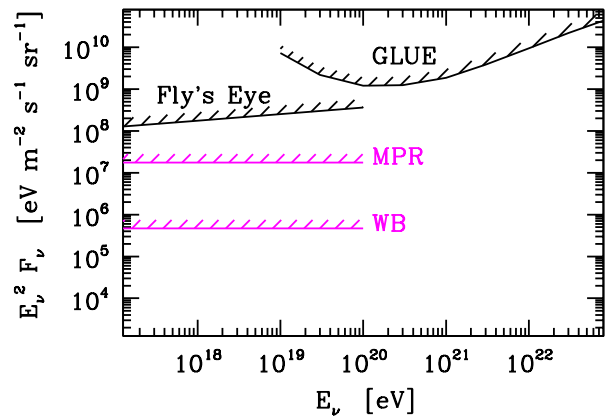}
\caption{\label{flux_upp_lim}
Upper limits on differential neutrino fluxes in the ultrahigh energy regime.
Shown are experimental upper limits 
on $F_{\nu_e}+F_{\bar\nu_e}$ from 
Fly's Eye~\cite{Baltrusaitis:1985mt} and 
on $\sum_{\alpha =e,\mu} (F_{\nu_\alpha }+F_{\bar\nu_\alpha })$ from 
the Gold\-stone lunar ultrahigh energy neutrino ex\-pe\-ri\-ment 
GLUE~\cite{Gorham:2001aj}, as well as theoretical upper limits on 
$F_{\nu_\mu}+F_{\bar\nu_\mu}$
from ``visible'' (``WB''~\cite{Waxman:1999yy,Bahcall:2001yr}) and 
``hidden'' (``MPR''~\cite{Mannheim:2001wp}) hadronic astrophysical sources.
}
\end{figure}

Presently unknown ingredients in the evaluation of the Z-burst spectra~(\ref{p-flux}) and  
(\ref{ph-flux}) are 
the differential fluxes $F_{\nu_i}$ of ultrahigh energy cosmic neutrinos (see e.\,g.   
Refs.~\cite{Protheroe:1999ei,Gandhi:2000kq,Learned:2000sw} for recent reviews). 
Present experimental upper limits on these fluxes are rather poor 
(cf. Fig.~\ref{flux_upp_lim} and Refs.~\cite{Seckel:2001,Yoshida:2001}).

What are the theoretical expectations for diffuse UHEC$\nu$ fluxes? 
More or less guaranteed are the so-called 
cosmogenic neutrinos which are produced when ultrahigh energy cosmic protons
scatter inelastically off the cosmic microwave background radiation~\cite{Greisen:1966jv,Zatsepin:1966jv}
in processes such as $p\gamma\to \Delta\to n\pi^+$,
where the produced pions subsequently 
decay~\cite{Beresinsky:1969qj,Beresinsky:1970}.
These fluxes (for recent estimates, see Refs.~\cite{Yoshida:1993pt,Protheroe:1996ft,Yoshida:1997ie,Engel:2001hd}) 
represent reasonable 
lower limits, 
but turn out to be insufficient for the
Z-burst scenario. 
Recently, theoretical upper limits on the ultrahigh energy cosmic neutrino flux have been given 
in Refs.~\cite{Waxman:1999yy,Mannheim:2001wp,Bahcall:2001yr}. Per construction, the upper
limit from ``visible'' hadronic astrophysical sources, i.\,e. from those sources which are transparent to ultrahigh 
energy cosmic protons
and neutrons, is of the order of the cosmogenic neutrino flux
and shown in Fig.~\ref{flux_upp_lim} (``WB''; cf.~Refs.~\cite{Waxman:1999yy,Bahcall:2001yr}). 
Also shown in this figure (``MPR'') is 
the much larger upper limit from ``hidden'' hadronic astrophysical sources, i.\,e. from those sources from which only 
photons and neutrinos can escape~\cite{Berezinsky:1979pd,Mannheim:2001wp}. 
Even larger fluxes at ultrahigh energies may arise if the hadronic astrophysical sources emit
photons only in the sub-MeV region -- thus evading the ``MPR'' bound in Fig.~\ref{flux_upp_lim} -- 
or if the neutrinos are produced via the decay of superheavy 
relic particles~\cite{Gelmini:2000ds,Crooks:2001jw,Ellis:1990iu,Ellis:1992nb,Gondolo:1993rn,Berezinsky:1997hy,%
Kuzmin:1997cm,Birkel:1998nx,Sarkar:2001se}, for which also the fragmentation function of the decay is of major 
interest~\cite{Berezinsky:1998ed,Berezinsky:2001up,Toldra:2002yz,Toldra:2002sn,Barbot:2002ep,Ibarra:2002rq},  
or topological defects~\cite{Bhattacharjee:2000qc,Berezinsky:2000az}.

In this situation of insufficient knowledge, we take the following approach concerning the flux of 
ultrahigh energy cosmic neutrinos, $F_{\nu_i}(E_{\nu_i},r)$. It is assumed to have the form 
\begin{equation}
F_{\nu_i}(E_{\nu_i},r)=F_{\nu_i}(E_{\nu_i},0)\,(1+z)^\alpha\,,
\end{equation} 
where $z$ is the redshift and where $\alpha$ characterizes the cosmological source evolution  
(see also Refs.~\cite{Yoshida:1997ie,Yoshida:1998it,Kalashev:2001sh}). The flux at zero redshift, 
$F_{\nu_i}(E_{\nu_i},0)\equiv F_{\nu_i}(E_{\nu_i})$,
is left open. 
For hadronic astrophysical sources it is expected to fall off power-like, $F_{\nu_i}(E_{\nu_i})\propto E_{\nu_i}^{-\gamma}$, 
$\gamma\gwig 1$, at high energies. Due to this fact and because of the strong resonance peaks in the $\nu_i\bar\nu_i$ annihilation 
cross section~(\ref{z-res-cs-roulet}) at the resonance energies~(\ref{eres}), the Z-burst rate will be only sensitive to 
the flux at the resonant energy of
the heaviest neutrino. Of course, the latter may be nearly degenerate with the other neutrino mass eigenstates, 
$m_{\nu_i}\approx m_\nu$, as it is the case for $m_{\nu_3}\,\gwig\, 0.1$~eV in a three flavour scenario. 
Correspondingly, our later fit to the UHECR data will be sensitive only to 
\begin{equation}
\label{fnures}
F_\nu^{\rm res} = \sum_i \left[ F_{\nu_i}(E_{\nu_i}^{\rm res})+F_{\bar\nu_i}(E_{\nu_i}^{\rm res})
\right] \,,
\end{equation}
where the sum extends over the number of mass eigenstates which are quasi-degenerate with the heaviest neutrino.
Note, finally, that, independently of the production
mechanism, neutrino oscillations result in a uniform $F_{\nu_i}$ mixture for the 
different mass eigenstates $i$.   

\begin{figure}
\vspace*{-5.0mm}
\hspace*{-2.cm}
\includegraphics[angle=90,width=12cm]
{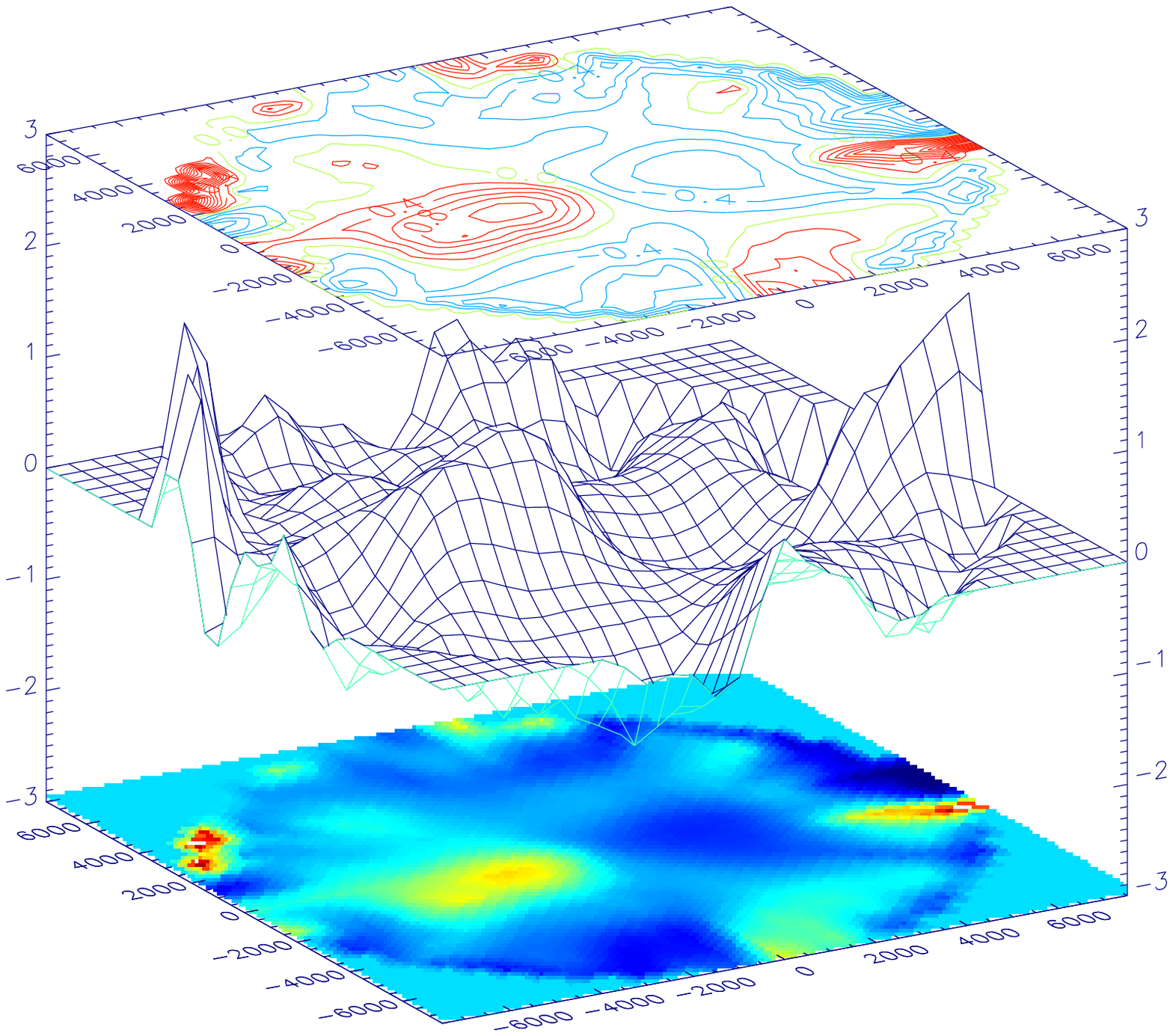}
\includegraphics[bbllx=20pt,bblly=225pt,bburx=570pt,bbury=608pt,width=8.65cm]
{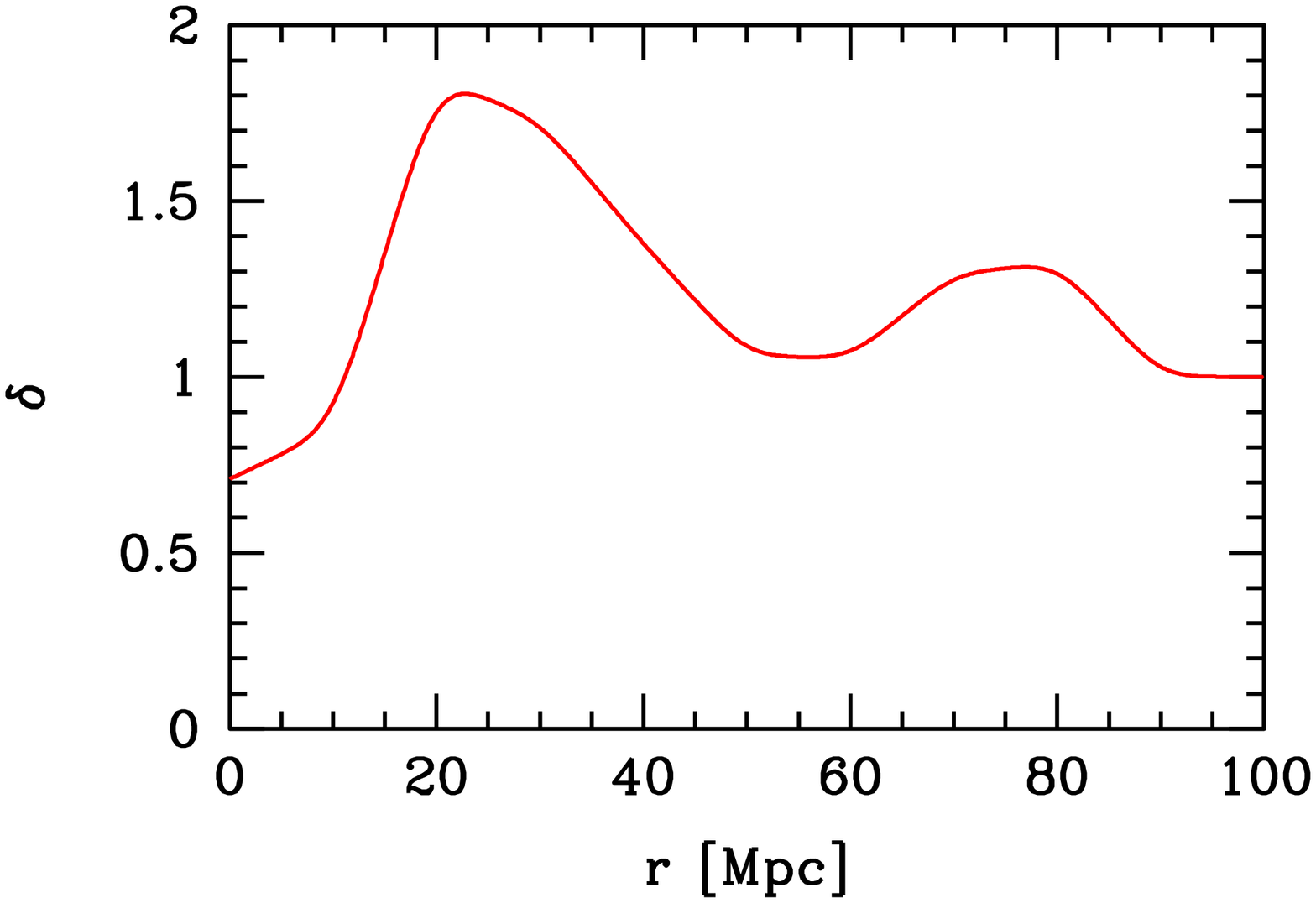}
\caption[...]{\label{dens-prof}
{\em Top:}
Mass density fluctuation field $\delta$ along the 
supergalactic plane as obtained from
peculiar velocity measurements~\cite{daCosta:1996nt}. Shown are contours
in intervals of $\delta =0.2$, surface maps on a grid of spacing 
$500$ km\,s$^{-1}$, corresponding to $5\,h^{-1}$ Mpc, with the height proportional to
$\delta$, and contrast maps. One recognizes some well-known structures in the
nearby volume such as the Great Attractor at supergalactic coordinates
(SGX $\sim -2000$ km\,s$^{-1}$, SGY $\sim -500$ km\,s$^{-1}$), 
the Perseus-Pisces complex 
(SGX $\sim 6000$ km\,s$^{-1}$, SGY $\sim -1000$ km\,s$^{-1}$), 
and the large void
(SGX $\sim 2500$ km\,s$^{-1}$, SGY $\sim 0$ km\,s$^{-1}$) in between.
{\em Bottom:} 
Mass density fluctuation field obtained from above data,  
averaged over all directions, for $h=0.71$. The overdensities at around 20 and 80 Mpc 
reflect the Great Attractor and the Perseus-Pisces complex, respectively. 
}
\end{figure}

\subsection{\label{nunumb}Neutrino number density}

The dependence of the relic neutrino number density $n_{\nu_i}$ on the distance $r$ 
is treated in the following way. 

The question is whether there is remarkable clustering of the relic neutrinos within
the local GZK zone of about 50 Mpc. It is known that the 
density distribution of relic neutrinos as hot dark matter follows the total mass 
distribution; however, with  less clustering~\cite{Ma:1998aw,Primack:2000iq}.
In fact, for $m_\nu\lwig 1$ eV, one expects pretty much that the neutrino number density
equals the big bang prediction~(\ref{standard_number_dens})~\cite{Primack:2001}. 
To take above facts into account, the shape of the $n_{\nu_i}(r)$ 
distribution is varied, for distances below 100 Mpc,  
between the standard cosmological homogeneous case~(\ref{standard_number_dens}) 
and that of 
the total mass distribution obtained from peculiar velocity 
measurements~\cite{daCosta:1996nt,Dekel99} (cf. Fig.~\ref{dens-prof} (top)). 
These peculiar measurements suggest relative overdensities of at most a factor $f_\nu =2\div 3$, 
depending on the grid spacing (cf. Fig.~\ref{dens-prof} (bottom)). 
A relative overdensity $f_\nu = 10^2\div 10^4$ in our neighbourhood, as it was assumed in earlier 
investigations of the Z-burst 
hypothesis~\cite{Fargion:1999ft,Weiler:1999sh,Waxman:1998yh,Yoshida:1998it,Blanco-Pillado:2000yb,McKellar:2001hk}, 
seems unlikely in view of these data. 
Our quantitative results turned out to be rather insensitive to the 
variations of the overdensities within the considered range.

For scales larger than 100 Mpc the relic neutrino 
density is taken according to the big bang cosmology prediction, 
$n_{\nu_i}=56\cdot (1+z)^3$ cm$^{-3}$. Possible uniform neutrino density enhancements due to 
eventual lepton asymmetries, as advocated in Ref.~\cite{Gelmini:1999qa}, are negligible in view of the recent, 
very stringent bounds on the neutrino degeneracies~\cite{Kneller:2001cd,Dolgov:2002ab}. 
In any case, such uniform enhancements change only the normalization of the Z-burst spectra and 
have no effect on their shape. Therefore, they could possibly milder the required UHEC$\nu$ flux, but
the value of the neutrino mass inferred from the Z-burst scenario is not affected by them.

\section{\label{sect:determination}Determination of the required neutrino mass and the necessary UHEC$\nu$ flux}

\subsection{\label{gen}Generalities}

\begin{figure}
\includegraphics[bbllx=20pt,bblly=221pt,bburx=570pt,bbury=608pt,width=7.9cm]
{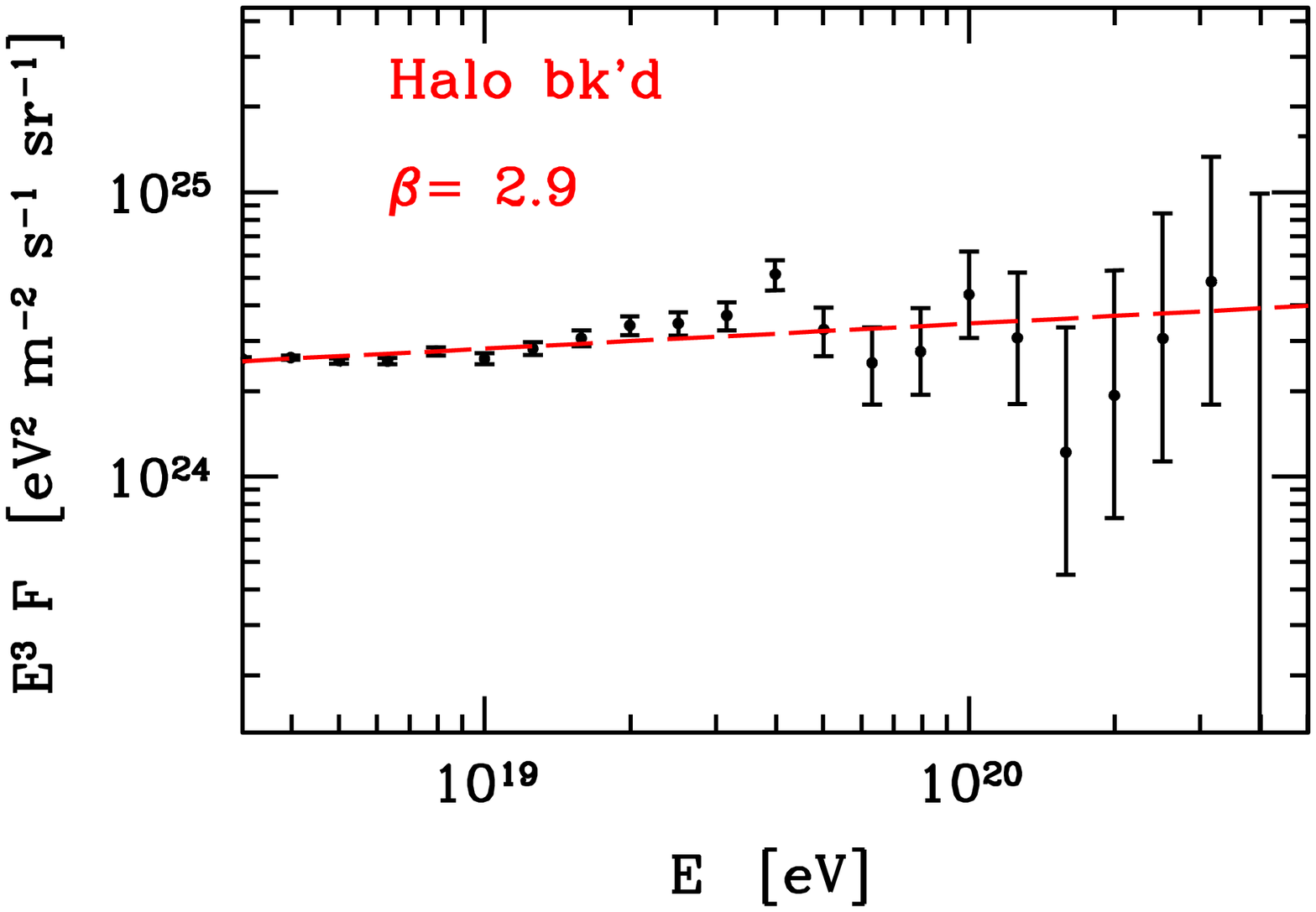}
\includegraphics[bbllx=20pt,bblly=221pt,bburx=570pt,bbury=608pt,width=7.9cm]
{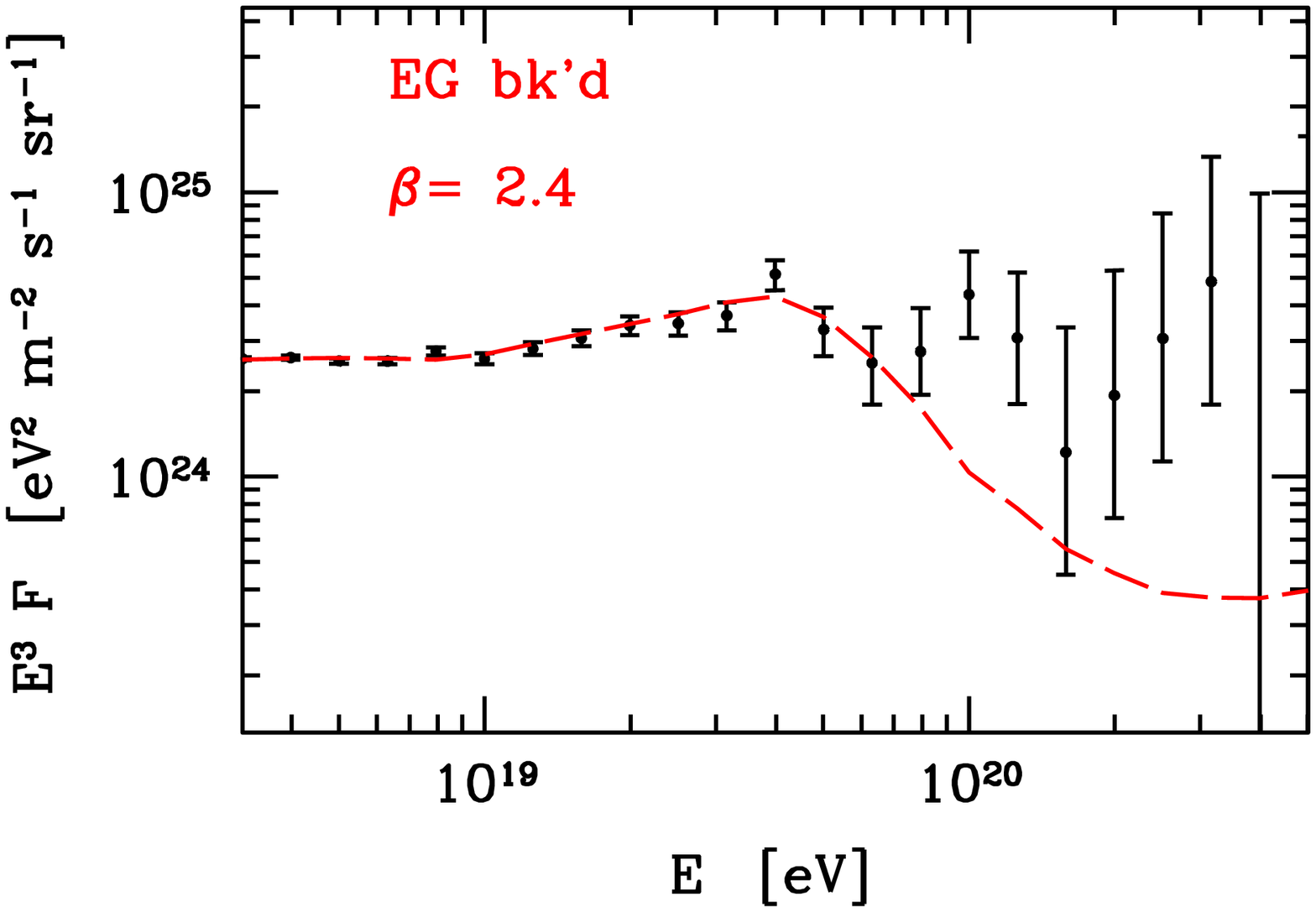}
\caption[...]{\label{fit_normal}
The available UHECR data with their error bars
and the best fits (long-dashed) from ordinary cosmic ray protons originating from our local neighbourhood
(``halo background''; {\em top}) and originating from diffuse extragalactic sources 
(``EG background''; {\em bottom}), respectively.
For the latter case, the bump at $4\cdot 10^{19}$~eV represents protons injected at
high energies and accumulated just above the GZK cutoff due to their energy
losses. The predicted fall-off for energies above $4\cdot 10^{19}$~eV can be observed.
}
\end{figure}

The predicted spectra of protons and photons from Z-bursts, (\ref{p-flux}) and (\ref{ph-flux}), 
can now be compared with the observed UHECR spectrum (cf. Fig.~\ref{fit_normal}). 
Our analysis includes published UHECR data
of AGASA~\cite{Takeda:1998ps}, Fly's Eye~\cite{Bird:yi,Bird:wp,Bird:1994uy}, 
Haverah Park~\cite{Lawrence:cc,Ave:2000nd}, 
and HiReS~\cite{Kieda00}, as well as unpublished one from the World Wide Web pages of 
the experiments on 17/03/01 (for a review, see Ref.~\cite{Nagano:2000ve}). 
Due to normalization difficulties we did not use the Yakutsk~\cite{Efimov91} results. 
We shall take into account the fact that above $4\cdot 10^{19}$~eV less than $50\,\%$ of 
the cosmic rays can be photons at the $95\,\%$ confidence level (C.L.)~\cite{Ave:2001xn} 
(see also Refs.~\cite{Ave:2000nd,Shinozaki:2001}).  

As usual, each logarithmic unit between 
$\log (E/\mbox{eV})=18$ and $\log (E/\mbox{eV})=26$
is divided into ten bins.
The predicted number of UHECR events in a bin is taken as
\begin{equation}
\label{flux}
 N(i)= {\mathcal E}
\int_{E_i}^{E_{i+1}}
{\rm d}E
\left[
F_{p|{\rm bkd}}(E  )
+ F_{p(+\gamma )|Z} (E )\right],
\end{equation}
where ${\mathcal E} \approx 8\,\cdot 10^{16}$ m$^2\cdot$\,s\,$\cdot$\,sr
is the total exposure (estimated from the highest energy
events and the corresponding fluxes) and where 
$E_i=10^{(18+i/10)}$~eV is the lower bound of the $i^{\rm th}$ energy bin. The first 
term in Eq.~(\ref{flux}), $F_{p|{\rm bkd}}$, corresponds to the diffuse
background of ordinary cosmic rays from unresolved astrophysical sources. 
Below the GZK cutoff, it should have the usual and experimentally observed 
power-law form~\cite{Nagano:2000ve}.   
The second term represents the sum of the proton and photon spectra, 
$F_{p|Z}+F_{\gamma |Z}$, Eqs.~(\ref{p-flux}) and (\ref{ph-flux}), from Z-bursts. 

The separation of the flux into two terms (one from the power-law   
background and one from the Z-burst) is physically well motivated. The
power-law part below the GZK cutoff is confirmed experimentally and, for
extragalactic sources, it should suffer from the GZK effect. In the Z-burst
scenario, cosmic rays are coming from another independent source
(Z-bursts), too. What we observe is the sum of the two. As the detailed
fits in the next section will show, the flux from Z-bursts is much smaller in the low energy
region than the flux of the power-law background. Correspondingly, the low energy part
of the spectrum (between $10^{18.5}$ and $10^{19.3}$~eV) has very little influence on the
Z-burst fit parameters, notably on the neutrino mass.

We shall study several possibilities for the background term $F_{p|{\rm bkd}}$. 

The first one is based on the assumption that the diffuse background of ordinary cosmic rays 
even above the GZK cutoff at $4\cdot 10^{19}$~eV consists of protons which are produced in 
our neighborhood -- within our galactic halo or at least within the GZK zone of about 50~Mpc. 
We shall refer to this possible cosmic ray background in the following as the ``halo background''. 
We should note, however, that there are no apparent suitable astrophysical sources within the
GZK distance to which the highest energy events point~\cite{Sigl:1994eg,Elbert:1995zv} -- a fact
which is difficult to reconcile with the halo background model. 

For the halo background, no GZK attenuation is included. The spectrum is assumed  
to have the usual power-law behavior which describes the data well for smaller
energies~\cite{Takeda:1998ps,Nagano:2000ve} (cf. Fig.~\ref{fit_normal} (top)), 
\begin{eqnarray}
\label{pow-law-halo}
\lefteqn{
{\mathcal E}\,F_{p|{\rm bkd}}(E;A,\beta ) = }
\\[1ex]\nonumber && \hspace{6ex}
\frac{A}{1\,{\rm eV}}\,
\left( \frac{E}{1\,{\rm eV}}\right)^{-\beta}
 \hspace{3ex} ({\rm Halo\ bk'd})
\,.
\end{eqnarray}
A similar background model, however with a fixed power-law index $\beta = 3.23$, as found by AGASA
to fit the data from $10^{17.6}\div 10^{19}$~eV (cf. Table V in Ref.~\cite{Nagano:2000ve}), 
has been exploited in a recent Z-burst simulation~\cite{Gelmini:2002xy}. 

In the second case,  
we assume that the diffuse background of ordinary cosmic rays comes from protons which originate from
uniformly distributed, extragalactic sources -- hence we name it ``extragalactic background''.
In view of the observed distribution of arrival directions of UHECRs, this assumption seems to be 
phenomenologically more realistic than the halo background model. Though there are some peculiarly
clustered events, the overall distribution at present statistics seems
to be practically uniform~\cite{Dubovsky:1998pu,Berezinsky:1999rp,MedinaTanco:1999gw}.

The extragalactic background suffers of course from GZK attenuation. Correspondingly, we take the above 
power-law~(\ref{pow-law-halo}),  
$A \cdot E^{-\beta}_p$, as an injection spectrum and take its modification due to the 
interactions with the CMB photons into account with the help of the proton propagation function 
$P_p(r,E_p;E)$, 
\begin{eqnarray}
\nonumber
\lefteqn{
{\mathcal E}\,F_{p|{\rm bkd}}(E;A,\beta ) = }
\\[1ex]\nonumber &&  \hspace{3ex}
 \int_0^\infty {\rm d}E_p \int_0^{R_{\rm max}} {\rm d}r 
\,\left(1+z(r)\right)^3
\\[1ex] \label{pow-law-eg} &&  \hspace{5ex}
\times\,
\frac{A}{1\,{\rm eV}}\,  \left( \frac{E_p}{1\,{\rm eV}}\right)^{-\beta}\,
\\[1ex]\nonumber &&   \hspace{5ex}
\times\,
(-)\frac{\partial P_p(r,E_p;E)}{\partial E}
 \hspace{3ex} 
({\rm EG\ bk'd})\,.
\end{eqnarray}
The predicted spectrum of the extragactic background protons shows an accumulation at around the GZK 
scale $4\cdot 10^{19}$~eV and a sharp drop beyond (see Fig.~\ref{fit_normal} (bottom)). 

As far as the Z-burst spectra entering our prediction~(\ref{flux}) are concerned, 
we proceed as follows. 
We shall mainly concentrate on the case where there is only one 
relevant neutrino mass scale $m_\nu$, either because there are three neutrino 
types with nearly degenerate neutrino masses, $m_{\nu_i}\approx m_\nu$, 
or there is one neutrino which is much heavier than the other ones such that
the contribution of the latter to the cosmic ray spectrum can be neglected since
the corresponding resonance energies are much larger and the UHEC$\nu$ fluxes
are expected to fall with increasing energy. Therefore, we fit only one neutrino mass
parameter $m_\nu$. In this case, the spectra~(\ref{p-flux}) and 
(\ref{ph-flux}) for protons and photons from Z-bursts can be written as
\begin{eqnarray}
\nonumber
\lefteqn{
{\mathcal E}\,F_{i|Z}(E;B,m_\nu ) = }
\\[1ex]\nonumber && 
B\, \int_0^\infty {\rm d}E_i \int_0^{R_{\rm max}} {\rm d}r 
\,\left(1+z(r)\right)^{3+\alpha}\,\delta_n (r) 
\\[1ex]\label{fit-zburst-spectra}  
 &&
\frac{4\,m_\nu}{M_Z^2}\,
{\mathcal Q_i}\left( y = \frac{4\,m_\nu\,E_i}{M_Z^2} \right)
\\[1ex]\nonumber
 && 
\times\,
(-)\frac{\partial P_i(r,E_i;E)}{\partial E}
,
 \hspace{6ex} i=p,\gamma\,,
\end{eqnarray}
where 
$\alpha$ is the cosmological evolution parameter,  
$\delta_n(r)$ is the mass density fluctuation field (cf. Fig.~\ref{dens-prof} (bottom)), normalized to one,  
and $\mathcal Q_i$ are the boosted momentum distributions from hadronic Z decay, normalized 
to $\langle N_{p+n}\rangle = 2.04$, for $i=p$, and to $\langle N_\gamma\rangle = 2\,\langle N_{\pi^0}\rangle +
\langle N_{\pi^\pm}\rangle = 37$, for $i=\gamma$. 
We are left here with two fit parameters, the mass $m_\nu$ of the heaviest neutrino and the 
overall normalization $B$, which may be expressed, on account of Eqs.~(\ref{eresfres}) and
(\ref{fnures}), in terms of the original quantities entering Eqs.~(\ref{p-flux}) and (\ref{ph-flux}), as 
\begin{eqnarray}
\label{B-uhecnu-fluxes}
\frac{B}{{\mathcal E}}= {\rm Br}(Z\to {\rm hadrons})\,R_{\rm max}\,\langle n_{\nu_i}\rangle_0\,
\langle \sigma_{\rm ann}\rangle\,E_\nu^{\rm res}\,F_\nu^{\rm res}
\,.
\end{eqnarray}
Note, that the neglection of finite width effects, $1/2\,\delta E_{\nu_i}^{\rm res}/E_{\nu_i}^{\rm res}=\Gamma_Z/M_Z=2.7\,\%$,
in our implementation~(\ref{fit-zburst-spectra})  
of the Z-burst spectra is perfectly adequate in view of the relative errors $\gwig 50$\,\% which we 
will find later from our fits.       

The expectation value for the number of events in a bin is given
by Eq.~(\ref{flux}). 
To determine the most probable value for $m_{\nu_j}$ we use the maximum 
likelihood method and minimize~\cite{Fodor:2001za} the 
$\chi^2(\beta,A,B,m_{\nu_j})$,
\begin{eqnarray} 
\label{chi}
\lefteqn{
\chi^2=}
\\[1ex] \nonumber &&
\sum_{\log\left(\frac{E_i}{{\rm eV}}\right)=18.5}^{\log\left(\frac{E_i}{\rm eV}\right)=26.0}
2\left[ N(i)-N_{\rm o}(i)+N_{\rm o}(i)
\ln\left( N_{\rm o}(i)/N(i)\right) \right],
\end{eqnarray}
where $N_{\rm o}(i)$ is the total number of observed events in the $i^{\rm th}$
bin. 
Since the Z-burst scenario results
in a quite small flux for lower energies, 
we take the lower bound just below the ``ankle'': $E_{\rm min}=10^{18.5}$~eV. Our results are
insensitive to the definition of the upper end (the flux is
extremely small there) for which we choose $\log (E_{\rm max}/\mbox{eV})=26$.
The uncertainties of the
measured energies are about 30\% which is one bin. 
By means of a Monte Carlo analysis, we take these uncertainties into account 
and include the corresponding variations in our final error estimates.

For comparison with recent work on the Z-burst scenario~\cite{Kalashev:2001sh}, 
we have  done also fits with no background component, $F_{p|{\rm bkd}}=0$ in 
Eq.~(\ref{flux}), and a larger lower end, $\log (E_{\rm min}/\mbox{eV})= 19.4\div 20.0$.
Such a scenario, with $E_{\rm min}\lwig E_{\rm GZK}\approx 4\cdot 10^{19}$~eV, was advocated
in Ref.~\cite{Kalashev:2001sh} as appropriate to explain all UHECR data above the GZK cutoff by the
Z-burst model and to attribute all UHECR data below the cutoff to ordinary astrophysical
extragalactic sources.

\subsection{Fit results}

Qualitatively, our analysis can be understood as follows.
In the Z-burst scenario a small relic neutrino mass needs a large incident neutrino energy 
$E_\nu^{\rm res}$~(\ref{eres}) in order to produce a Z. Large $E_\nu^{\rm res}$ results in 
a large Lorentz boost, thus large $E_p$ resp. $E_\gamma$. In this way the {\em shape} of the
detected energy ($E$) spectrum determines the mass of the relic neutrino. 
The sum of the necessary UHEC$\nu$ fluxes  $F_\nu^{\rm res}$, on the other hand, is determined by the 
over-all {\em normalization} $B$.

Our fitting procedure involves four parameters: $\beta,A,B$ and 
$m_\nu$. The minimum of the $\chi^2(\beta,A,B,m_\nu)$ function is 
$\chi^2_{\rm min}$ at $m_{\nu\, {\rm min}}$ which is
the most probable value for the mass, whereas
the 1\,$\sigma$ (68\%) confidence interval for $m_\nu$ 
is determined by 
\begin{equation}
\chi^2(\beta',A',B',m_\nu)\equiv \chi^2_o(m_\nu)=\chi^2_{\rm min}+1
\,.
\end{equation}
Here $\beta'$, $A'$, $B'$ are defined in such a way that the 
$\chi^2$ function is minimized in $\beta,A$
and $B$, at fixed $m_\nu$.

\begin{figure}
\includegraphics[bbllx=20pt,bblly=221pt,bburx=570pt,bbury=608pt,width=7.9cm]
{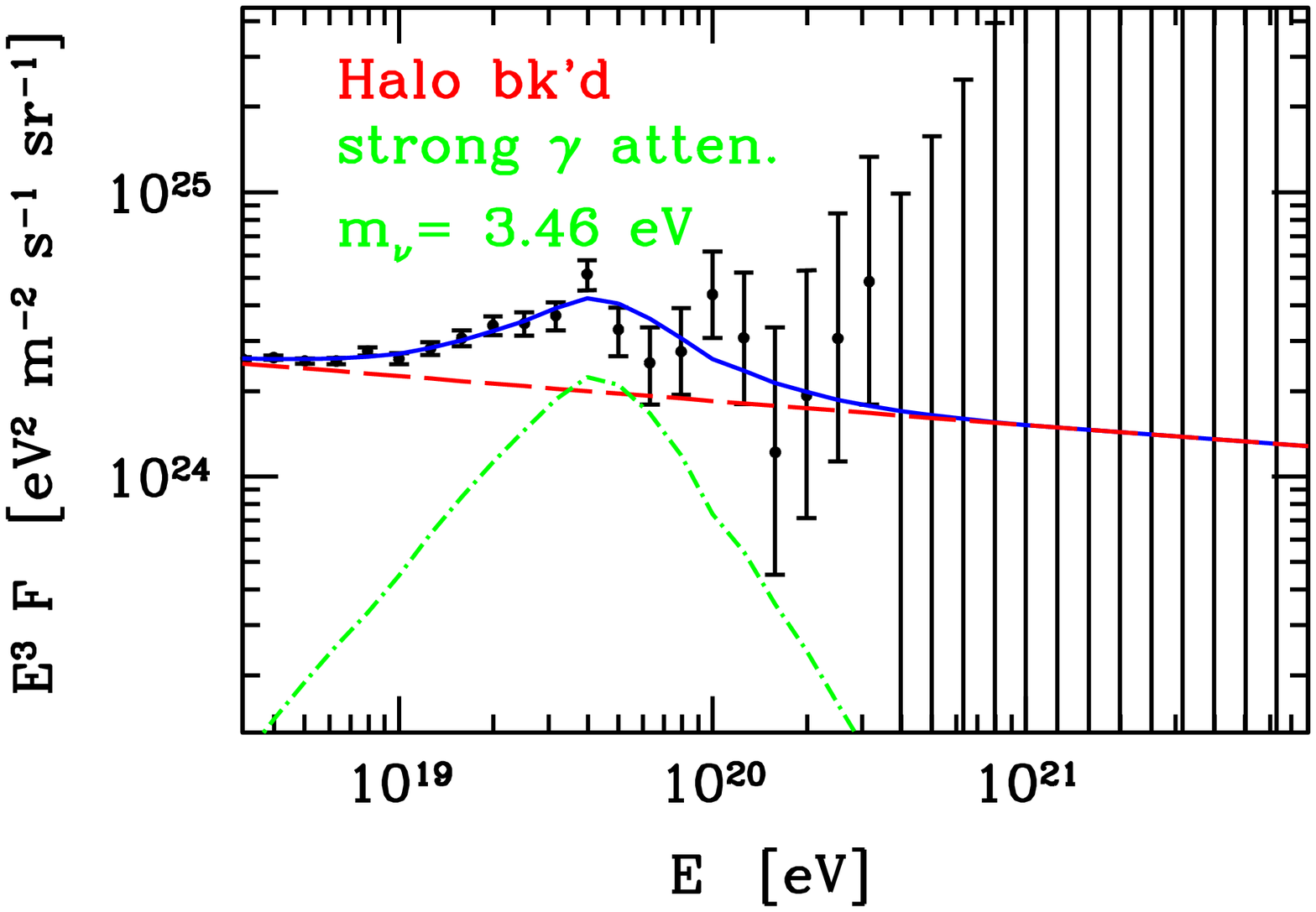}
\includegraphics[bbllx=20pt,bblly=221pt,bburx=570pt,bbury=608pt,width=7.9cm]
{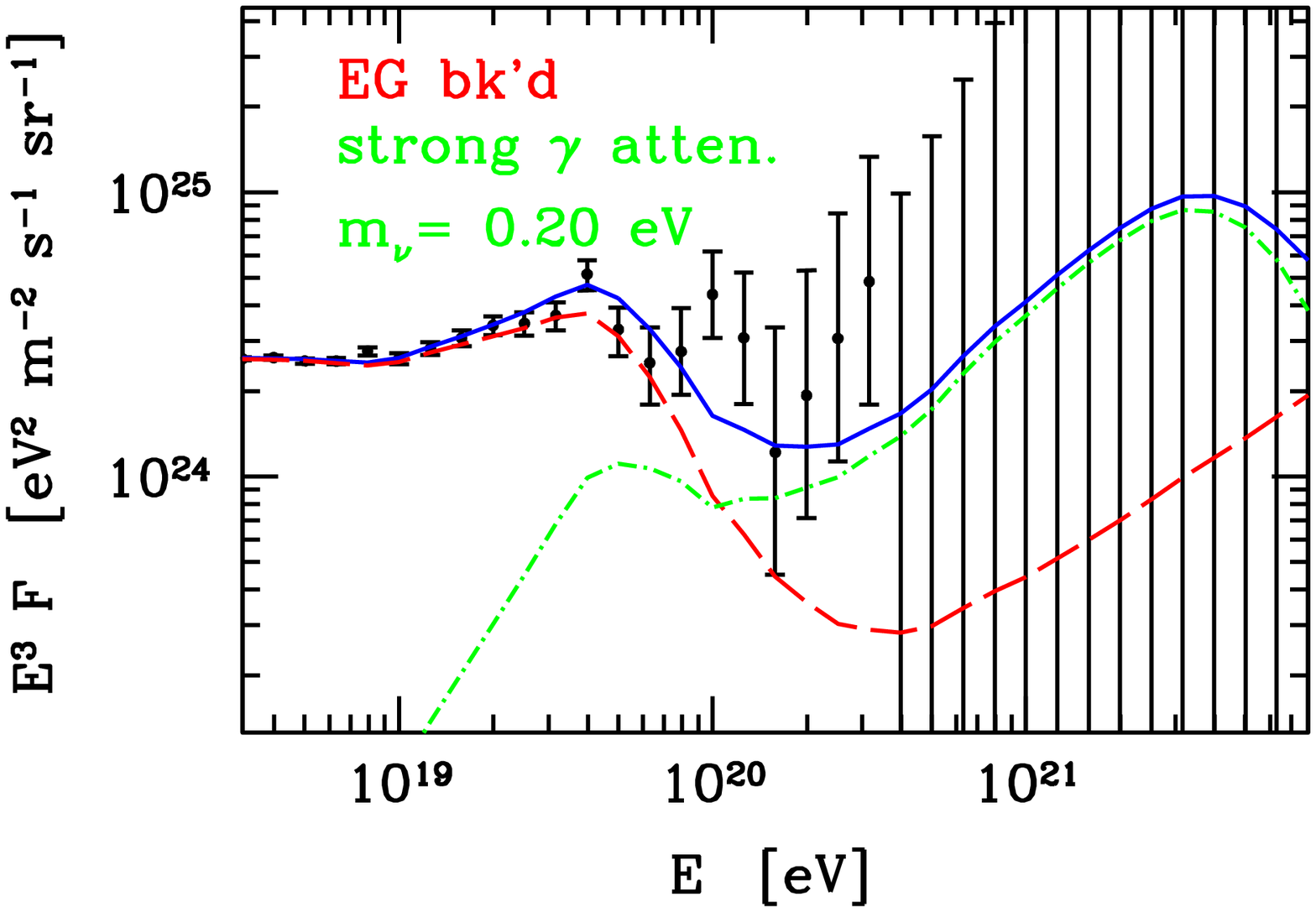}
\includegraphics[bbllx=20pt,bblly=221pt,bburx=570pt,bbury=608pt,width=7.9cm]
{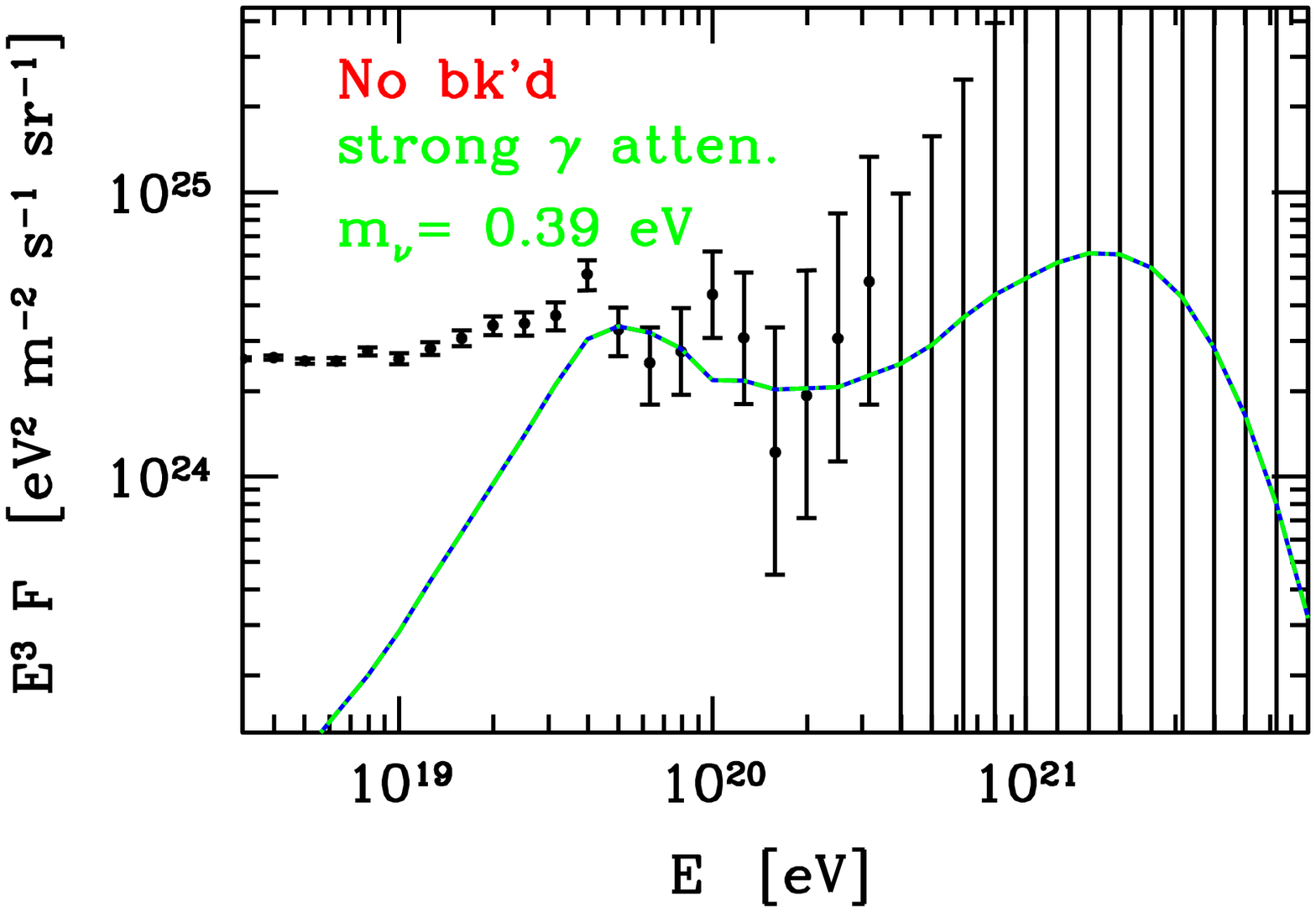}
\caption[...]{\label{fit_nu}
The available UHECR data with their error bars
and the best fits from Z-bursts, for 
a strong UHE$\gamma$ attenuation such that
the Z-burst photons can be neglected ($\alpha =0, h=0.71, \Omega_M= 0.3,\Omega_\Lambda =0.7,z_{\rm max}=2$).
{\em Top:} 
Best fit for the case of a halo background (solid line). The bump 
around $4\cdot 10^{19}$~eV is mainly due to the Z-burst protons (dash-dotted), whereas 
the 
almost horizontal contribution (long-dashed) is the first, power-law-like 
term of Eq.~(\ref{flux}). 
{\em Middle:} 
The case of an ``extragalactic'' UHECR background.
The first bump at $4\cdot 10^{19}$~eV represents protons produced at
high energies and accumulated just above the GZK cutoff due to their energy
losses. The bump at $2\cdot 10^{21}$~eV is a remnant of the Z-burst
energy. The long-dashed line shows the contribution of the power-law-like
spectrum with the GZK effect included. 
{\em Bottom:} 
The case of no UHECR background above $\log (E_{\rm min}/\mbox{eV})= 19.7$. 
}
\end{figure}

\begin{table}
\caption{\label{fit_noph}Results of fits, for 
a strong UHE$\gamma$ attenuation ($h=0.71,\Omega_M= 0.3,\Omega_\Lambda =0.7,z_{\rm max}=2$). 
{\em Top:} Assuming a halo UHECR background according to Eq.~(\ref{pow-law-halo}).
{\em Middle:} Assuming an extragalactic UHECR background according to Eq.~(\ref{pow-law-eg}).
{\em Bottom:} Assuming no UHECR background above $E_{\rm min}$, for different values of the 
lower end $E_{\rm min}$ of the fit ($\alpha = 0$).}
\begin{ruledtabular}
\begin{tabular}{|r||c|c|c|c|c|}
\multicolumn{6}{|c|}{Halo UHECR background + strong UHE$\gamma$ attenuation}\\ \hline
$\alpha$ & $m_\nu$ [eV] & $\chi^2_{\rm min}$ & $A$ & $B$ & $\beta$\\ \hline
$-3$&   $4.46^{+2.22(4.80)}_{-1.64(2.88)}$&     $15.90$&        $1.46\cdot 10^{43}$&    $1049$& $3.110$\\
$0$&    $3.46^{+1.73(4.03)}_{-1.34(2.32)}$&     $15.64$&        $1.62\cdot 10^{43}$&    $770$&  $3.111$\\
$3$&    $2.51^{+1.45(3.30)}_{-1.05(1.80)}$&     $15.53$&        $1.65\cdot 10^{43}$&    $551$&  $3.111$\\
\hline
\multicolumn{6}{|c|}{EG UHECR background + strong UHE$\gamma$ attenuation}\\ \hline
$\alpha$ & $m_\nu$ [eV] & $\chi^2_{\rm min}$ & $A$ & $B$ & $\beta$\\ \hline
$-3$&   $0.20^{+0.20(0.63)}_{-0.11(0.18)}$&     $25.82$&        $5.00\cdot 10^{31}$&    $150$&  $2.465$\\
$0$&    $0.20^{+0.19(0.61)}_{-0.12(0.18)}$&     $26.41$&        $5.98\cdot 10^{31}$&    $144$&  $2.466$\\
$3$&    $0.20^{+0.19(0.59)}_{-0.11(0.17)}$&     $26.89$&        $7.23\cdot 10^{31}$&    $142$&  $2.467$\\
\hline
\multicolumn{6}{|c|}{No UHECR background + strong UHE$\gamma$ attenuation}\\ \hline
$\log (E_{\rm min}/\mbox{eV})$ & $m_\nu$ [eV] & $\chi^2_{\rm min}$ & $A$ & $B$ & $\beta$\\ \hline
$19.4$& $2.28^{+0.64(1.46)}_{-0.58(1.06)}$&     $21.81$&        $-$&    $1251$& $-$\\
$19.5$& $1.31^{+0.63(1.44)}_{-0.53(0.80)}$&     $16.01$&        $-$&    $846$&  $-$\\
$19.6$& $0.85^{+0.67(1.62)}_{-0.31(0.55)}$&     $14.80$&        $-$&    $670$&  $-$\\
$19.7$& $0.40^{+0.32(0.87)}_{-0.16(0.27)}$&     $8.03$& $-$&    $445$&  $-$\\
$19.8$& $0.42^{+0.41(1.25)}_{-0.18(0.29)}$&     $7.99$& $-$&    $460$&  $-$\\
$19.9$& $0.76^{+1.06(2.50)}_{-0.39(0.58)}$&     $5.52$& $-$&    $733$&  $-$\\
$20.0$& $1.77^{+1.49(3.47)}_{-1.01(1.47)}$&     $2.68$& $-$&    $2021$& $-$\\
\end{tabular}
\end{ruledtabular}
\end{table}

As already mentioned, presently there is no evidence that the observed highest energy cosmic
rays are photons. Let us start therefore with the assumption (cf. Ref.~\cite{Fodor:2001qy}) 
that the ultrahigh energy photons from 
Z-bursts can be neglected in the fit in comparison to the protons. This is certainly true
for a sufficiently large universal radio background, e.\,g. on the level of the maximal one 
estimated in Ref.~\cite{Protheroe:1996si} and/or for a sufficiently strong extragalactic 
magnetic field ${\mathcal O}(10^{-9})$~G. 
We shall refer to this scenario in the following as ``strong'' UHE$\gamma$ attenuation.  
Our best fits to the observed data from this scenario~(cf. Table~\ref{fit_noph}) can be seen in 
Fig.~\ref{fit_nu}, for evolution parameter $\alpha=0$. 
We find a neutrino mass of
\massnophhalo
for the case that the UHECR background protons
are of halo type~(\ref{pow-law-halo}), \massnopheg   
if they are of 
extragalactic type~(\ref{pow-law-eg}), and 
\massnophno
if there are are no background protons above 
$10^{19.7}$~eV, respectively. The first numbers are
the 1\,$\sigma$, the numbers in the brackets are the
2\,$\sigma$ errors. This gives an absolute lower bound on the mass of the 
heaviest neutrino of 
\lowbdnopheg
at the 95\% C.L., which is comparable to the one obtained from the atmospheric
mass splitting in a three flavour scenario, Eq.~(\ref{lim_low_atm}).

\begin{figure}
\begin{center}
\includegraphics[bbllx=20pt,bblly=221pt,bburx=570pt,bbury=608pt,width=8.6cm]
{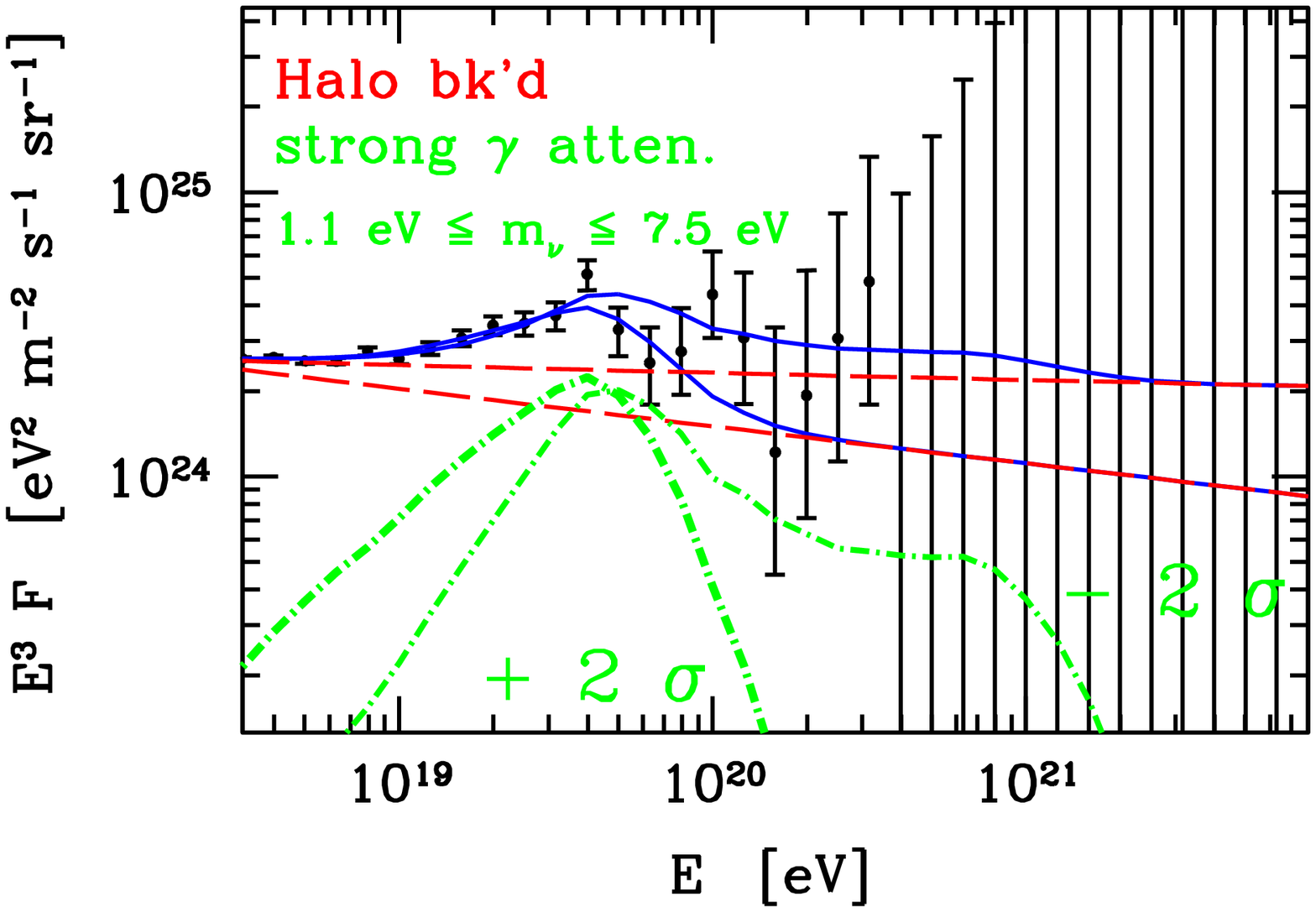}
\includegraphics[bbllx=20pt,bblly=221pt,bburx=570pt,bbury=608pt,width=8.6cm]
{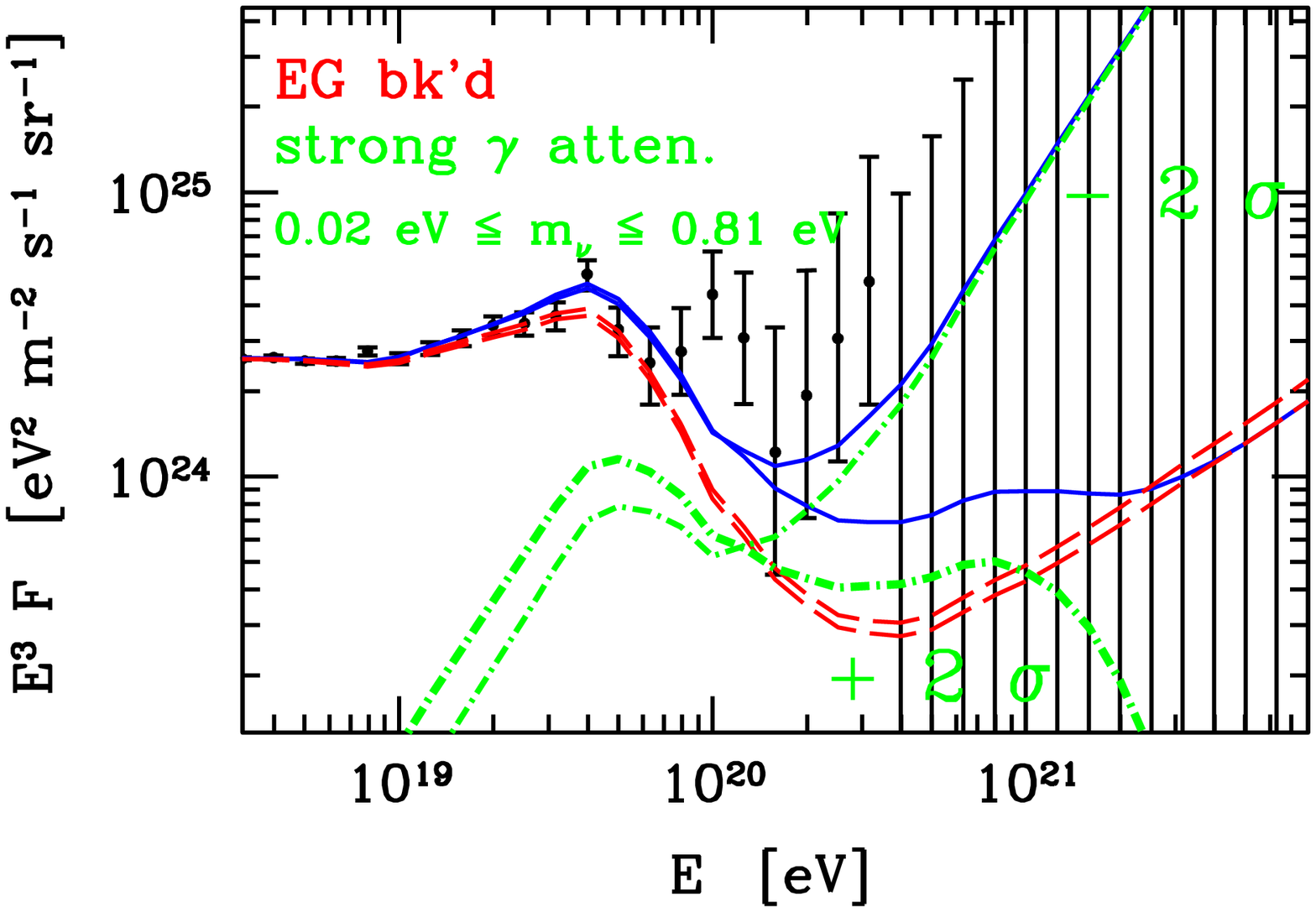}
\includegraphics[bbllx=20pt,bblly=221pt,bburx=570pt,bbury=608pt,width=8.6cm]
{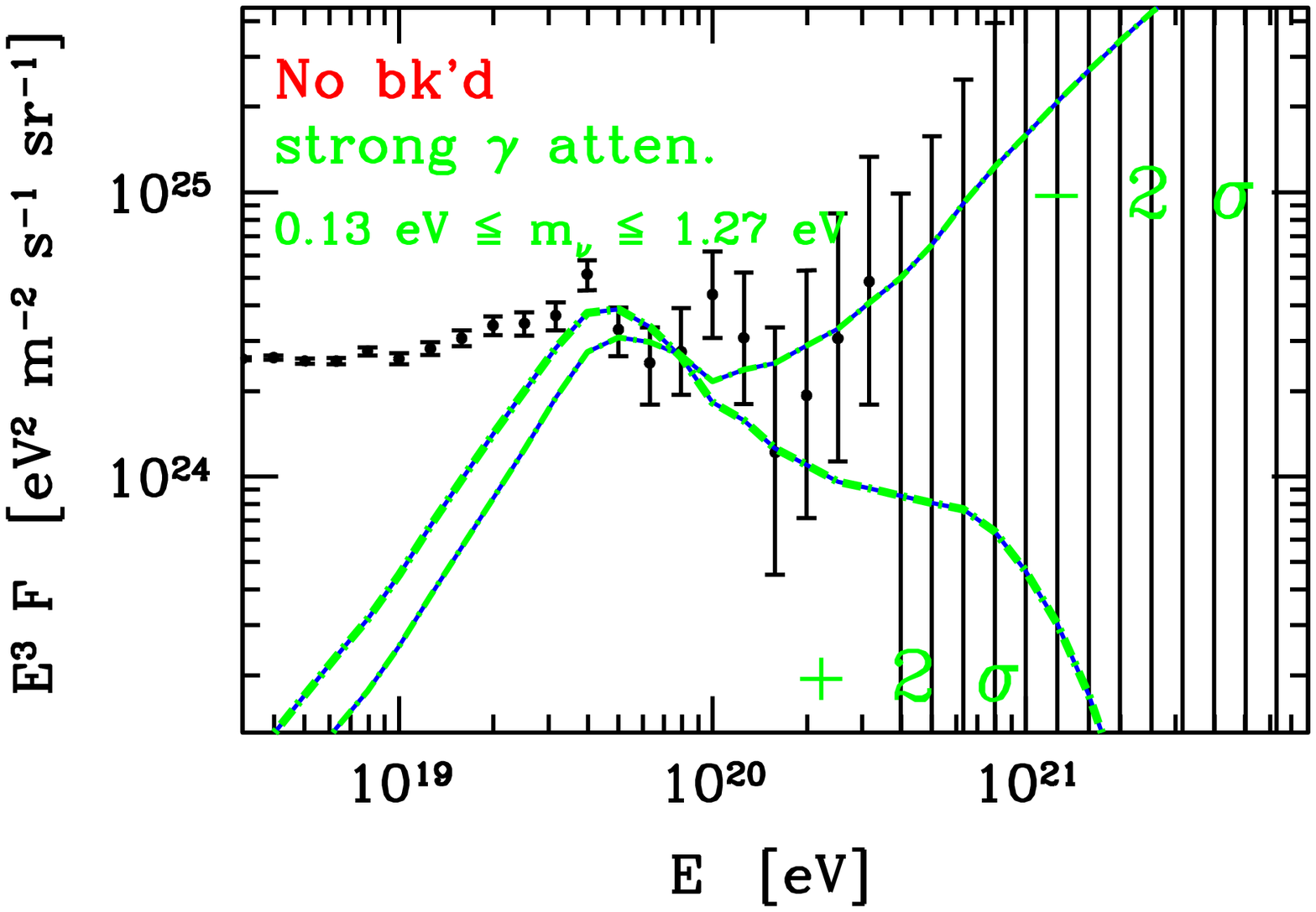}
\caption[...]{\label{fit_errors}
The available UHECR data with their error bars
and the $\chi^2_{\rm min}\pm 4$ fits (solid lines) from Z-bursts, 
for a halo ({\em top}), an extragalactic ({\em middle}), and no ({\em bottom})  
UHECR background (long-dashed lines), and 
a strong UHE$\gamma$ attenuation  such that
the Z-burst photons can be neglected 
and only the Z-burst protons (dash-dotted lines) have to be taken into account in the fit 
($\alpha =0, h=0.71, \Omega_M= 0.3,\Omega_\Lambda =0.7,z_{\rm max}=2$).
}
\end{center}
\end{figure}

The surprisingly small uncertainties are based on the 
$\chi^2$ analysis described in Section~\ref{gen}.  
The inclusion of the already mentioned 30\,\% uncertainties in the observed energies
by a Monte Carlo analysis increases the error bars by about 10\,\%. 
Note, that the relative errors in the extragalactic and in the no 
background cases are of the same order. This shows that the smallness of these
errors does not originate from the low energy part of the background component.  

The fits are rather good:
for 21 non-vanishing bins and 4 fitted parameters they can be as low as 
\chiminnophhalo
and 
\chiminnopheg 
in the halo and the extralactic background case, respectively,  
whereas in the no background case, for $E_{\rm min}=10^{19.7}$~eV, we have 
9 bins with 2 fitted parameters and a 
\chiminnophno
see Table~\ref{fit_noph}.  
In the latter case, however, which was advocated strongly in Ref.~\cite{Kalashev:2001sh}, a remarkable
dependence of the fitted mass on the value of $E_{\rm min}$ is observed. 
The $\pm\, 2\ \sigma$ fits are shown in
Fig.~\ref{fit_errors}. As it should be, the spread is small in the region where 
there are data, whereas it can be  quite large in the presently unexplored ultrahigh energy regime.
Finally, let us mention that, 
both in the ``halo'' as well as in the extragalactic background
cases, the $\chi^2$ fits {\em without} a Z-burst component (cf. Fig.~\ref{fit_normal}) are far worse: we find 
\chiminnozhalo\ 
in the former, and \chiminnozeg\ in the latter, for 21 non-vanishing bins and 2 fitted parameters. 
This finding suggests that the power-law background terms alone cannot describe the data.

\begin{figure}
\begin{center}
\includegraphics[bbllx=20pt,bblly=221pt,bburx=570pt,bbury=608pt,width=8.6cm]
{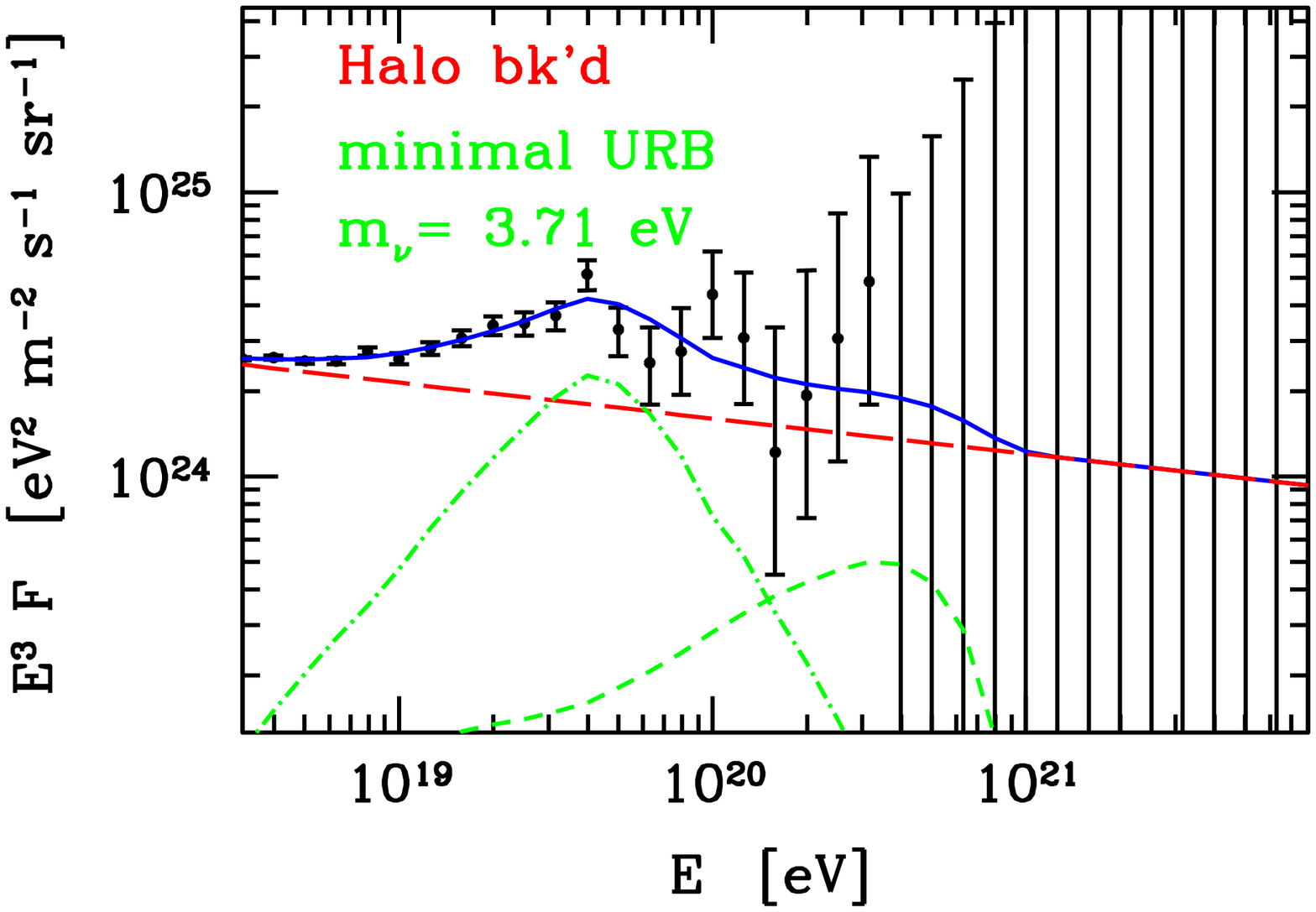}
\includegraphics[bbllx=20pt,bblly=221pt,bburx=570pt,bbury=608pt,width=8.6cm]
{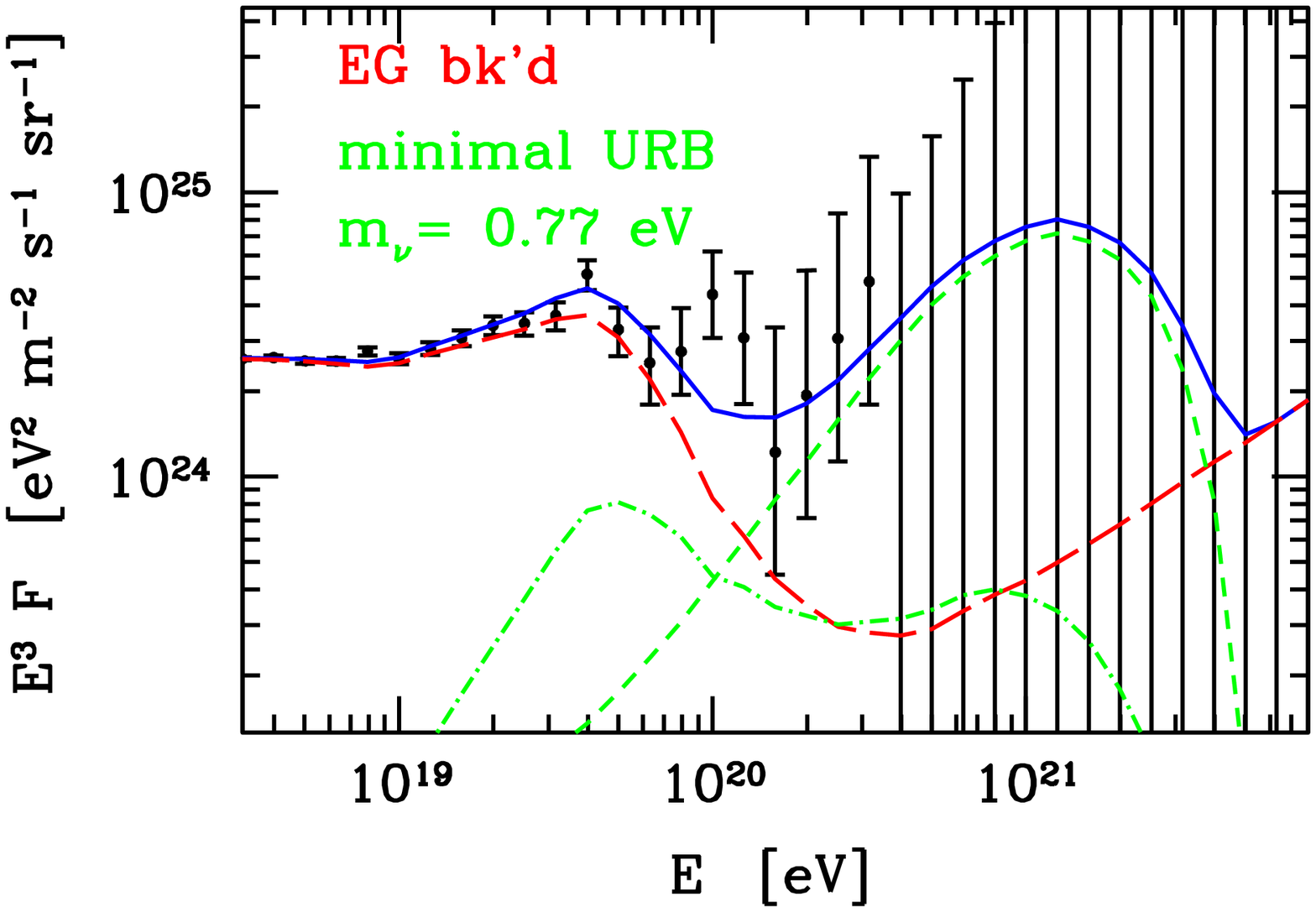}
\includegraphics[bbllx=20pt,bblly=221pt,bburx=570pt,bbury=608pt,width=8.6cm]
{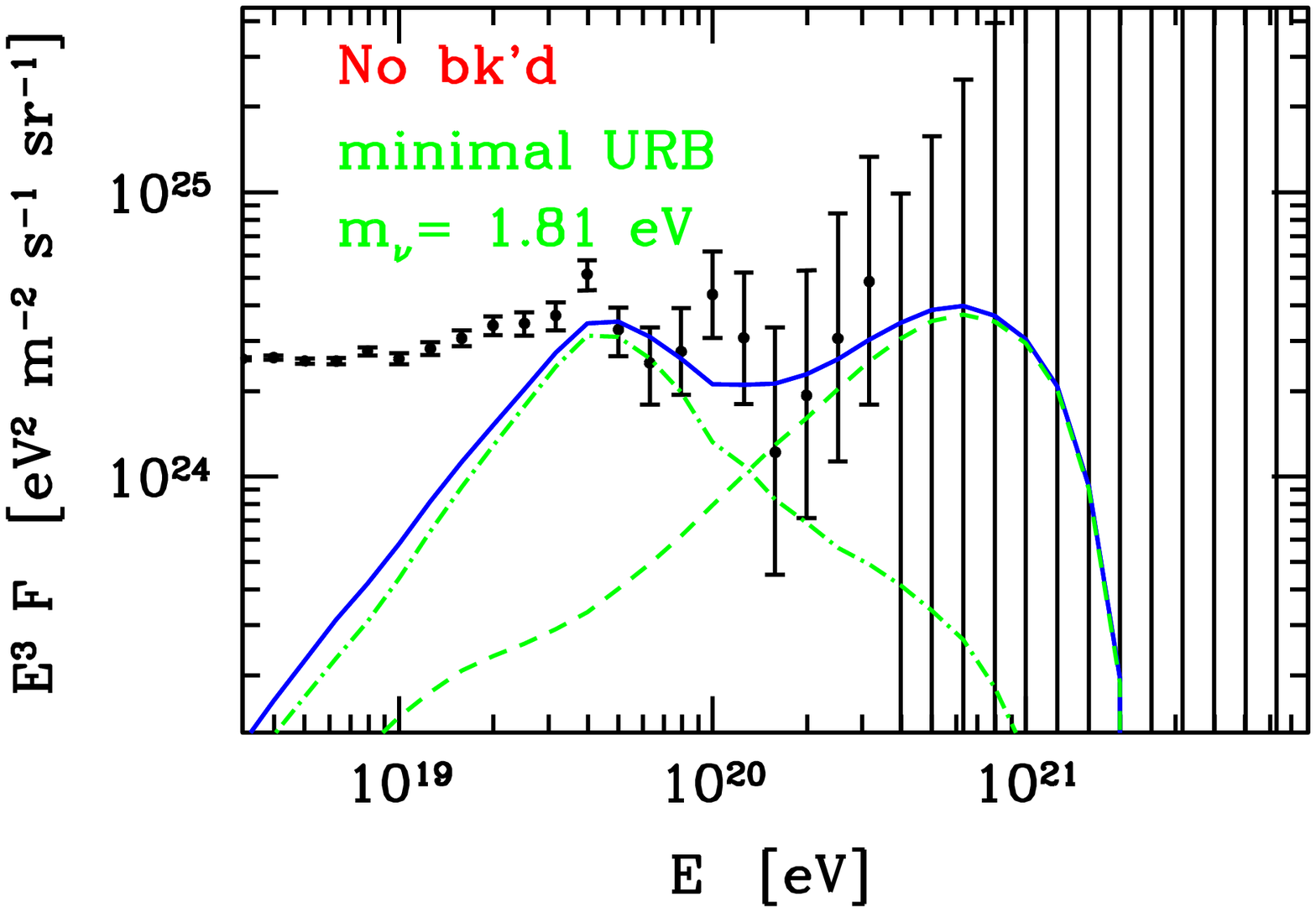}
\caption[...]{\label{fit_nu_minurb}
The available UHECR data with their error bars
and the best fits from Z-bursts, 
for an energy attenuation  
of photons as in Ref.~\cite{Lee:1998fp} and a vanishing EGMF  
($\alpha=0,h=0.71,\Omega_M= 0.3,\Omega_\Lambda =0.7,z_{\rm max}=2$).
{\em Top:} 
Best fit for the case of a halo UHECR background (solid line), corresponding to the sum of the 
background protons (long-dashed), the Z-burst protons (dash-dotted)
and the Z-burst photons (short-dashed).
{\em Middle:} 
The case of an extragalactic UHECR background.
{\em Bottom:} 
The case of no UHECR background above $\log (E_{\rm min}/\mbox{eV})= 19.7$. 
}
\end{center}
\end{figure}

In addition to the case of strong UHE$\gamma$ attenuation, corresponding to a large universal 
radio background or a large extragalactic magnetic field, let us consider now the case of a 
``minimal'' URB. Here, we assume a vanishing EGMF and exploit, as in 
Ref.~\cite{Lee:1998fp}, a universal radio background on the level of the one from Ref.~\cite{Clark:1970} 
(cf. Section~\ref{prop} and Fig.~\ref{ph_mfp} (top)). 
The corresponding fits, for our three background scenarios, can be seen in Fig.~\ref{fit_nu_minurb} 
and Table~\ref{fit_minurb}. Note, that our best fits are still compatible with the already mentioned upper limits on
the photon fraction of the observed ultrahigh energy cosmic rays~\cite{Ave:2000nd,Ave:2001xn,Shinozaki:2001}.
We observe that the value of the neutrino mass found in both the halo background scenario, 
with 
\massleehalo
 as well as in the extragalactic background scenario, 
with 
\massleeeg
 are compatible, at about the 2~$\sigma$ level, 
with the corresponding values found in the case of strong UHE$\gamma$ attenuation. As before, 
in the ``no'' background scenario, we find a strong dependence of the fitted mass on the value of $E_{\rm min}$.

\begin{table}
\caption{\label{fit_minurb}Results of fits, for 
a photon attenuation as in Ref.~\cite{Lee:1998fp} (``minimal'' URB) and vanishing EGMF 
($h=0.71,\Omega_M= 0.3,\Omega_\Lambda =0.7,z_{\rm max}=2$).  
{\em Top:} Assuming a halo UHECR background according to Eq.~(\ref{pow-law-halo}).
{\em Middle:} Assuming an extragalactic UHECR background according to Eq.~(\ref{pow-law-eg}).
{\em Bottom:} Assuming no UHECR background above $E_{\rm min}$, for different values of the 
lower end $E_{\rm min}$ of the fit ($\alpha = 0$).
}
\begin{ruledtabular}
\begin{tabular}{|r||c|c|c|c|c|}
\multicolumn{6}{|c|}{Halo UHECR background + minimal URB}\\ \hline
$\alpha$ & $m_\nu$ [eV] & $\chi^2_{\rm min}$ & $A$ & $B$ & $\beta$\\ \hline
$-3$&   $4.46^{+1.78(3.99)}_{-1.36(2.39)}$&     $15.69$&        $4.76\cdot 10^{43}$&    $1060$& $3.146$\\
$0$&    $3.71^{+1.40(3.27)}_{-1.12(1.96)}$&     $15.36$&        $1.05\cdot 10^{44}$&    $827$&  $3.166$\\
$3$&    $3.01^{+1.13(2.62)}_{-0.92(1.56)}$&     $15.20$&        $1.55\cdot 10^{44}$&    $642$&  $3.171$\\
\hline
\multicolumn{6}{|c|}{EG UHECR background + minimal URB}\\ \hline
$\alpha$ & $m_\nu$ [eV] & $\chi^2_{\rm min}$  & $A$ & $B$ & $\beta$\\ \hline
$-3$&   $0.79^{+0.49(1.40)}_{-0.31(0.53)}$&     $24.19$&        $5.73\cdot 10^{31}$&    $125$&  $2.466$\\
$0$&    $0.77^{+0.48(1.36)}_{-0.30(0.51)}$&     $24.52$&        $6.44\cdot 10^{31}$&    $120$&  $2.467$\\
$3$&    $0.79^{+0.45(1.32)}_{-0.31(0.53)}$&     $24.77$&        $7.55\cdot 10^{31}$&    $122$&  $2.468$\\
\hline
\multicolumn{6}{|c|}{No UHECR background + minimal URB}\\ \hline
$\log (E_{\rm min}/\mbox{eV})$ & $m_\nu$ [eV] & $\chi^2_{\rm min}$ & $A$ & $B$ & $\beta$\\ \hline
$19.4$& $3.08^{+0.78(1.68)}_{-0.58(1.05)}$&     $14.43$&        $-$&    $1380$& $-$\\
$19.5$& $2.56^{+0.76(1.68)}_{-0.55(0.97)}$&     $11.88$&        $-$&    $1111$& $-$\\
$19.6$& $2.51^{+0.84(1.87)}_{-0.60(1.05)}$&     $11.86$&        $-$&    $1079$& $-$\\
$19.7$& $1.81^{+0.75(1.76)}_{-0.51(0.86)}$&     $8.02$& $-$&    $680$&  $-$\\
$19.8$& $2.04^{+0.95(2.17)}_{-0.67(1.11)}$&     $7.74$& $-$&    $812$&  $-$\\
$19.9$& $3.01^{+1.34(3.03)}_{-1.06(1.78)}$&     $4.93$& $-$&    $1699$& $-$\\
$20.0$& $4.67^{+2.12(4.60)}_{-1.74(2.95)}$&     $2.33$& $-$&    $4881$& $-$\\
\end{tabular}
\end{ruledtabular}
\end{table}

\begin{figure}[b]
\includegraphics[bbllx=20pt,bblly=221pt,bburx=570pt,bbury=608pt,width=8.65cm]
{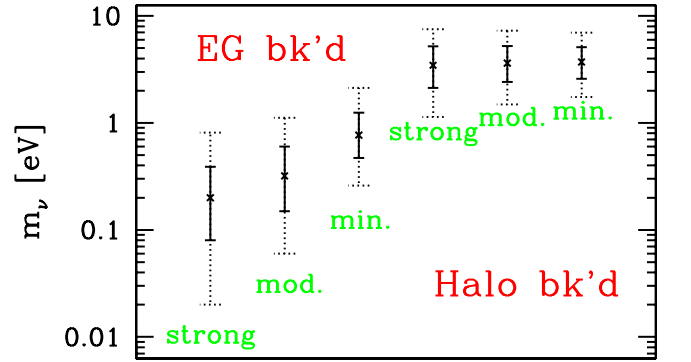}
\caption[...]{\label{mass_res}
Summary of the masses of the heaviest neutrino required in the Z-burst scenario, with
their 1\,$\sigma$ (solid) and 2\,$\sigma$ (dotted) error bars, for the case of an extragalactic and a halo  
background of ordinary cosmic rays and   
for various assumptions about the diffuse extragalactic photon background in the radio band 
($\alpha =0, h=0.71, \Omega_M= 0.3,\Omega_\Lambda =0.7,z_{\rm max}=2$).
From left: strong $\gamma$ attenuation, moderate and minimal URB.
}
\end{figure}

Besides the no UHECR background scenario, the Z-burst determination of the neutrino mass seems reasonably robust.  
Its robustness with regard to changes of presently unknown quantities within their
anticipated variations is further illustrated in Tables~\ref{fit_modurb} and \ref{fit_hubble} and in Fig.~\ref{mass_res}. 
For a wide range of cosmological source evolution ($\alpha =-3\div 3$), 
Hubble parameters $h=0.61\div 0.9$, $\Omega_M$, $\Omega_\Lambda$, $z_{\rm max}=2\div 5$, for variations of the possible relic neutrino
overdensity in our GZK zone within the limits discussed in Section~\ref{nunumb},  
and for different assumptions about the diffuse extragalactic photon background,   
the results remain within the above error bars. The main uncertainties concerning the 
central values originate from the different assumptions about the background of ordinary 
cosmic rays. In the case that the ordinary cosmic rays above $10^{18.5}$~eV are protons and 
originate from a region within the GZK zone of about 50~Mpc (``halo''), the required mass of the 
heaviest neutrino seems to lie between
\massrangehaloonesig
at the 68\,\% C.L. ($\alpha\leq 0$, cf. below), if we take into account the variations between the
minimal and moderate URB cases and the strong UHE$\gamma$ attenuation case (cf. Fig.~\ref{mass_res}).
Note, that a value of $m_\nu = 0.07$~eV, as studied recently in Ref.~\cite{Gelmini:2002xy} in
a simulation of the Z-burst scenario in the context of a halo background model, seems to lie
more than 3\,$\sigma$ away from our best fit values. 
The much more plausible assumption that the ordinary cosmic rays above $10^{18.5}$ are 
protons of extragalactic origin leads to a required neutrino mass of
\massrangeegonesig\ 
at the 68\,\% C.L. ($\alpha\leq 0$) (cf. Fig.~\ref{mass_res}).

\begin{table}
\caption{\label{fit_modurb}Results of fits, for 
a photon attenuation as in Fig.~\ref{ph_mfp} (bottom), short-dashed line (``moderate'' URB) and vanishing EGMF 
($h=0.71,\Omega_M= 0.3,\Omega_\Lambda =0.7,z_{\rm max}=2$).  
{\em Top:} Assuming a halo UHECR background according to Eq.~(\ref{pow-law-halo}).
{\em Middle:} Assuming an extragalactic UHECR background according to Eq.~(\ref{pow-law-eg}).
{\em Bottom:} Assuming no UHECR background above $E_{\rm min}$, for different values of the 
lower end $E_{\rm min}$ of the fit ($\alpha = 0$).
}
\begin{ruledtabular}
\begin{tabular}{|r||c|c|c|c|c|}
\multicolumn{6}{|c|}{Halo UHECR background + moderate URB}\\ \hline
$\alpha$ & $m_\nu$ [eV] & $\chi^2_{\rm min}$ & $A$ & $B$ & $\beta$\\ \hline
$-3$&   $4.35^{+1.97(4.31)}_{-1.45(2.57)}$&     $15.69$&        $2.16\cdot 10^{43}$&    $848$&  $3.112$\\
$0$&    $3.62^{+1.61(3.66)}_{-1.20(2.13)}$&     $15.44$&        $2.38\cdot 10^{43}$&    $662$&  $3.113$\\
$3$&    $2.94^{+1.30(3.00)}_{-1.01(1.72)}$&     $15.26$&        $4.91\cdot 10^{43}$&    $514$&  $3.149$\\
\hline
\multicolumn{6}{|c|}{EG UHECR background + moderate URB}\\ \hline
$\alpha$ & $m_\nu$ [eV] & $\chi^2_{\rm min}$  & $A$ & $B$ & $\beta$\\ \hline
$-3$&   $0.32^{+0.29(0.84)}_{-0.17(0.27)}$&     $21.15$&        $4.95\cdot 10^{31}$&    $60$&   $2.465$\\
$0$&    $0.32^{+0.28(0.80)}_{-0.17(0.26)}$&     $21.29$&        $5.40\cdot 10^{31}$&    $58$&   $2.466$\\
$3$&    $0.31^{+0.28(0.78)}_{-0.17(0.26)}$&     $21.37$&        $5.94\cdot 10^{31}$&    $56$&   $2.466$\\
\hline
\multicolumn{6}{|c|}{No UHECR background + moderate URB}\\ \hline
$\log (E_{\rm min}/\mbox{eV})$ & $m_\nu$ [eV] & $\chi^2_{\rm min}$ & $A$ & $B$ & $\beta$\\ \hline
$19.4$& $2.81^{+0.74(1.65)}_{-0.59(1.05)}$&     $15.01$&        $-$&    $1087$& $-$\\
$19.5$& $2.18^{+0.73(1.65)}_{-0.52(0.92)}$&     $11.99$&        $-$&    $820$&  $-$\\
$19.6$& $2.04^{+0.81(1.84)}_{-0.56(0.97)}$&     $11.82$&        $-$&    $755$&  $-$\\
$19.7$& $1.12^{+0.59(1.50)}_{-0.40(0.64)}$&     $5.53$& $-$&    $369$&  $-$\\
$19.8$& $1.04^{+0.75(1.92)}_{-0.44(0.69)}$&     $5.49$& $-$&    $341$&  $-$\\
$19.9$& $1.66^{+1.31(3.03)}_{-0.81(1.23)}$&     $4.26$& $-$&    $652$&  $-$\\
$20.0$& $2.81^{+2.15(5.03)}_{-1.44(2.24)}$&     $2.75$& $-$&    $1728$& $-$\\
\end{tabular}
\end{ruledtabular}
\end{table}

\begin{table}
\caption{\label{fit_hubble}Results of fits, for 
strong and minimal photon attenuation (cf. Tables~\ref{fit_noph} and \ref{fit_minurb}, respectively), for extremal values
of the (reduced) Hubble constant ($\alpha = 0,\Omega_M= 0.3,\Omega_\Lambda =0.7,z_{\rm max}=2$).}
\begin{ruledtabular}
\begin{tabular}{|l||c|c|c|c|c|}
\multicolumn{6}{|c|}{EG UHECR background + strong UHE$\gamma$ attenuation}\\ \hline
$h$ & $m_\nu$ [eV] & $\chi^2_{\rm min}$ & $A$ & $B$ & $\beta$\\ \hline
$0.61$& $0.19^{+0.19(0.62)}_{-0.12(0.17)}$&     $28.27$&        $1.47\cdot 10^{31}$&    $126$&  $2.452$\\
$0.90$& $0.21^{+0.20(0.64)}_{-0.11(0.18)}$&     $24.20$&        $8.93\cdot 10^{32}$&    $168$&  $2.529$\\
\hline
\multicolumn{6}{|c|}{EG UHECR background + minimal URB}\\ \hline
$h$ & $m_\nu$ [eV] & $\chi^2_{\rm min}$ & $A$ & $B$ & $\beta$\\ \hline
$0.61$& $0.64^{+0.42(1.23)}_{-0.25(0.44)}$&     $26.28$&        $1.46\cdot 10^{31}$&    $97$&   $2.452$\\
$0.90$& $1.02^{+0.64(1.70)}_{-0.39(0.65)}$&     $22.39$&        $1.20\cdot 10^{33}$&    $162$&  $2.531$\\
\hline
\multicolumn{6}{|c|}{EG UHECR background + moderate URB}\\ \hline
$h$ & $m_\nu$ [eV] & $\chi^2_{\rm min}$ & $A$ & $B$ & $\beta$\\ \hline
$0.61$& $0.26^{+0.25(0.71)}_{-0.14(0.22)}$&     $22.91$&        $8.03\cdot 10^{30}$&    $52$&   $2.435$\\
$0.90$& $0.40^{+0.37(1.03)}_{-0.20(0.32)}$&     $19.53$&        $8.41\cdot 10^{32}$&    $64$&   $2.529$\\
\end{tabular}
\end{ruledtabular}
\end{table}

We performed a Monte Carlo analysis studying higher statistics. In the near 
future, the Pierre Auger Observatory~\cite{Guerard:1999ym,Bertou:2000ip} will provide a ten times 
higher statistics, which 
reduces the error bars in the neutrino mass to about one third of their 
present values.

\begin{figure}
\includegraphics[bbllx=20pt,bblly=221pt,bburx=570pt,bbury=608pt,width=8.65cm]
{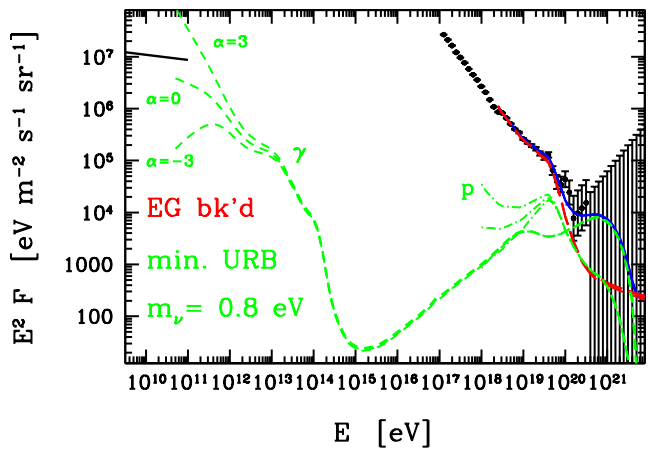}
\caption[...]{\label{fit_egret}
The available UHECR data with their error bars
and the best fit from Z-bursts, 
for various cosmological evolution parameters $\alpha$ and 
an energy attenuation of photons as in Fig.~\ref{ph_mfp} (bottom) exploiting  
a ``minimal'' URB  
($h=0.71, \Omega_M= 0.3,\Omega_\Lambda =0.7,z_{\rm max}=2$).  
Also shown is the diffuse $\gamma$ background in the energy range 
between 30 MeV and 100 GeV as measured by EGRET~\cite{Sreekumar:1998} (solid). 
}
\end{figure}

Let us consider next in more detail the $\gamma$ ray spectra from Z-bursts, notably
in the $\sim 100$~GeV region. As illustrated in Fig.~\ref{fit_egret}, 
the EGRET measurements of the diffuse $\gamma$ background in the energy range 
between 30~MeV and 100~GeV~\cite{Sreekumar:1998} put non-trivial constraints
on the evolution parameter $\alpha$. 
Whereas different values of $\alpha$ lead to very similar spectra in the ultrahigh energy region 
-- which determines the neutrino mass -- they
are easily discriminated in $\lwig$~TeV photons. Only evolution parameters $\alpha\lwig 0$ seem to 
be compatible with the EGRET measurements (cf. Fig.~\ref{fit_egret}), quite independently of different 
assumptions about the URB.      
These numerical findings are in fairly good agreement with other recent simulations~\cite{Kalashev:2001sh}.

\begin{figure}
\includegraphics[bbllx=20pt,bblly=221pt,bburx=570pt,bbury=608pt,width=8.65cm]
{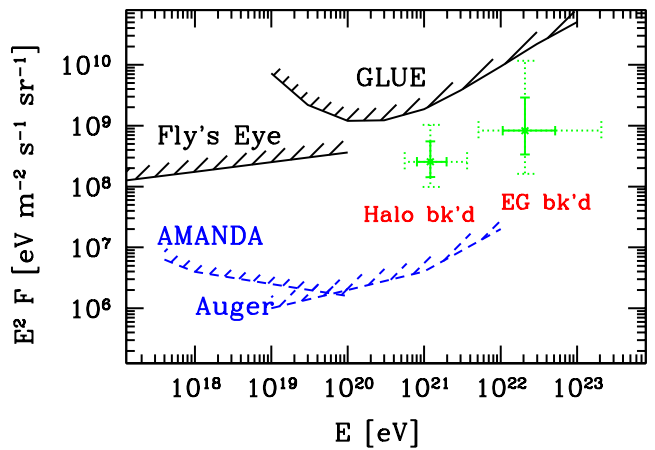}
\caption[...]{\label{eflux}
Neutrino fluxes, 
$F = \frac{1}{3} \sum_{i=1}^3 ( F_{\nu_i}+F_{\bar\nu_i})$, required 
by the Z-burst hypothesis for the case of a halo and an extragalactic background of 
ordinary cosmic rays, respectively ($\alpha =0, h=0.71, \Omega_M= 0.3,\Omega_\Lambda =0.7,z_{\rm max}=2$).
Shown are the necessary fluxes obtained from the fit results of Table~\ref{fit_noph} 
for the case of a strong UHE$\gamma$ attenuation.
The horizontal errors indicate the 1\,$\sigma$ (solid) and 2\,$\sigma$ (dotted) uncertainties of the
mass determination and the vertical errors include also the uncertainty
of the Hubble expansion rate.
Also shown are upper limits from Fly's Eye~\cite{Baltrusaitis:1985mt} on $F_{\nu_e}+F_{\bar\nu_e}$ and the 
Gold\-stone lunar ultrahigh energy neutrino ex\-pe\-ri\-ment 
GLUE~\cite{Gorham:2001aj} on $\sum_{\alpha = e,\mu} ( F_{\nu_\alpha}+F_{\bar\nu_\alpha})$, 
as well as projected sen\-si\-tivi\-ties of AMAN\-DA~\cite{Barwick00,Hundertmark:2001} on $F_{\nu_\mu}+F_{\bar\nu_\mu}$ and  
Auger~\cite{Yoshida:1998it,Capelle98} on $F_{\nu_e}+F_{\bar\nu_e}$. The sensitiviy of RICE is comparable to 
the one of Auger~\cite{Seckel:2001}.}
\end{figure}

The necessary UHEC$\nu$ flux at the resonant energy $E_\nu^{\rm res}$ is obtained via Eq.~(\ref{B-uhecnu-fluxes}) from
the fitted overall normalization $B$, 
\begin{equation}
\label{uhecnu-fluxes-B}
E_\nu^{\rm res}\,F_\nu^{\rm res} =
8.6\cdot 10^{-16}\,\frac{1}{{\rm m}^2\,{\rm s}\,{\rm sr}}
\,\frac{D_{H_0}}{R_{\rm max}}\,h\,B
\,,
\end{equation}
where $D_{H_0}=c/H_0=3\cdot 10^3\,h^{-1}$ Mpc is the present size of the universe and 
where, on account of Eqs.~(\ref{z-H}) and (\ref{H-Omega}), the ratio of the maximally considered 
distance of Z production, $R_{\rm max}$, and the size of the universe is given by
\begin{equation}
\frac{R_{\rm max}}{D_{H_0}} =
\int\limits_0^{{ z_{\rm max}}} {\rm d}z\,
\frac{1}{(1+z)\,
\sqrt{\Omega_{M}\,(1+z)^3 
+ \Omega_{\Lambda}}}\,.
\end{equation}
This ratio is of order one; it equals $0.73$ for $\Omega_M=0.3$, $\Omega_\Lambda =0.7$ and $z_{\rm max}=2$. 
The required fluxes are summarized in Fig.~\ref{eflux}, together with some
existing upper limits and projected sensitivities of 
pre\-sent, near future and future observational projects. 
They appear to be well below present upper limits and are within the expected sensitivity of 
AMANDA~\cite{Barwick00,Hundertmark:2001},  RICE~\cite{Seckel:2001}, and Auger~\cite{Capelle98}.  
Clearly, these fluxes are higher than the ones advocated in Ref.~\cite{Yoshida:1998it} based
on local neutrino overdensities $f_\nu =10^2\div 10^4$ on scales of about $5$ Mpc. However, since we also took into account
a background from ordinary cosmic rays, our normalization of the Z-burst component is different and correspondingly 
our fluxes are somewhat less than a factor of $f_\nu$ higher. They are consistent with the ones found in 
Refs.~\cite{Kalashev:2001sh,Gelmini:2002xy}. As far as power-law extrapolations of the fluxes, 
$F_\nu\propto E^{-\gamma}$, below the 
resonance energy are concerned, we find, in agreement with Refs.~\cite{Yoshida:1998it,Blanco-Pillado:2000yb,Kalashev:2001sh}, 
that indices $\gamma \gwig 1.5$ are excluded by the Fly's Eye limits~\cite{Baltrusaitis:1985mt} (cf. Fig.~\ref{eflux}). 

The required neutrino flux, for power-law extrapolations with index $\gamma \gwig 1$ to lower energies, 
is larger than the theoretical upper limit from ``hidden'' hadronic astrophysical sources, from which only 
photons and neutrinos can escape (cf. Fig.~\ref{flux_upp_lim} (``MPR'')).
From an analysis of the assumptions behind this limit~\citep{Mannheim:2001wp} 
one finds that one has to invoke hidden astrophysical sources whose photons somehow
don't show up in the diffuse $\gamma$ ray background measured by EGRET~\cite{Ringwald:2001mx,Kalashev:2001sh}. 
In general, astrophysical sources for the required UHE neutrinos should be distributed with
$\alpha <0$, accelerate protons up to energies $\gwig\, 10^{23}$~eV, be opaque to primary
nucleons and emit secondary photons only in the sub-MeV region. 
It is an interesting question whether such challenging conditions can be realized in BL Lac objects, a 
class of active galactic nuclei for which some evidence of zero or negative cosmological evolution has been 
found (see Ref.~\cite{Caccianiga:2001} and references therein) and which were recently discussed as 
possible sources of the highest energy cosmic rays~\cite{Tinyakov:2001nr}. 
Alternatively, one may invoke top-down scenarios~\cite{Bhattacharjee:2000qc} for
the sources of the highest energy cosmic neutrinos such as unstable superheavy relic
particle decays~\cite{Gelmini:2000ds,Crooks:2001jw,Ellis:1990iu,Ellis:1992nb,Gondolo:1993rn,Berezinsky:1997hy,%
Kuzmin:1997cm,Birkel:1998nx,Sarkar:2001se} or topological defect decays~\cite{Berezinsky:2000az}.

It should be stressed that, besides the neutrino mass, the UHEC$\nu$ flux at the resonance 
energy is one of the most robust predictions of the Z-burst scenario which can be verified
or falsified in the near future. In this connection it is worthwhile to recall (cf. Section~\ref{nunumb}) 
that the current limits on neutrino degeneracies~\cite{Kneller:2001cd,Dolgov:2002ab} do not
allow a remarkable uniform enhancement due to lepton asymmetries which otherwise would be 
welcome for a relaxation of the huge flux requirement~\cite{Gelmini:1999qa}. 

\section{\label{sect:discussion}Discussion and Conclusions}

We have presented a detailed comparison of the predicted spectra from 
Z-bu\-rsts -- resulting from the resonant annihilation of ultrahigh energy cosmic neutrinos
with relic neutrinos --  
with the observed ultrahigh energy cosmic ray spectrum, extending our earlier 
study~\cite{Fodor:2001qy}. The mass of the heaviest relic neutrino turned out to have to lie in the
range \massrangehaloonesig 
on the 1~$\sigma$ level, if 
the background of ordinary cosmic rays above $10^{18.5}$~eV consists of protons and 
originates from a region within a distance of about 50~Mpc. 
In the phenomenologically most plausible case that the ordinary cosmic rays above $10^{18.5}$ are 
protons of extragalactic origin one is lead to a required neutrino mass in the range
\massrangeegonesig\ at the 68\,\% C.L.. 
We have also analysed a case where there is no background of ordinary cosmic rays
above, say, $10^{19.7}$~eV. Here, we find \massrangenoonesig.
The above neutrino mass ranges include variations in presently 
unknown quantities,  like the amount of neutrino clustering, the universal radio background, and the
extragalactic magnetic field, within their anticipated uncertainties. 
If it turns out that the highest energy cosmic rays are extragalactic protons rather than photons  
one has to invoke a large universal radio background and/or large extragalactic magnetic field
to suppress the photons from Z-bursts. In this case and for our extragalactic background scenario, 
the neutrino mass range narrows down to \massrangeegstrongonesig, with a best fit value of \massnophegbestfit.  

It is remarkable, that the mass ranges required in the Z-burst scenario coincide nearly perfectly
with the present knowledge about the mass of the heaviest neutrino from oscillations and tritium
$\beta$ decay from Eqs.~(\ref{lim_low_atm}) and (\ref{lim_comb_osc_beta}), $0.04\ {\rm eV}\,\lwig\, m_{\nu_3}\,\lwig\,2.5$~eV, 
in a three flavour or, from Eqs.~(\ref{lim_low_lsnd}) and (\ref{lim_comb_osc_beta_lsnd}), 
$0.4\ {\rm eV}\,\lwig\, m_{\nu_4}\,\lwig\,3.8$~eV, in a four flavour scenario.     
Our values, in the extragalactic background scenario, are 
still compatible with a hierarchical neutrino mass spectrum, with the largest mass suggested by 
the atmospheric neutrino oscillation, $m_{\nu_3}\approx\sqrt{\triangle m^2_{\rm atm}}\approx~0.04\div 0.08$~eV. 
However, our best fit values point more into the direction of a quasi-degenerate scenario, 
$m_{\nu_i}\approx~m_\nu\gg~\sqrt{\triangle m^2_{\rm atm}}$. It is amusing to observe that the recently
reported evidence for neutrinoless double beta decay, with   
$0.11\ {\rm eV}\leq\langle m_\nu\rangle\leq 0.56$~eV,
if true, would point also into the same direction~\cite{Klapdor-Kleingrothaus:2002ke}. 

The above neutrino masses are in a range which can be explored by near-future laboratory experiments, 
like the KATRIN tritium $\beta$ decay experiment~\cite{Osipowicz:2001sq}, with a projected sensitivity of 
$m_\beta\gwig\, 0.3$~eV within $6\div 7$~years, or the $0\nu 2\beta$ 
decay experiments NEMO-3~\cite{Marquet:2000eq} and CUORE~\cite{Alessandrello:2002sj}, 
which aim at $\langle m_\nu\rangle \gwig\, 0.1$~eV, and their next-generation follow-up projects
GENIUS~\cite{Klapdor-Kleingrothaus:1998td} and EXO~\cite{Danilov:2000pp}, with $\langle m_\nu\rangle\gwig\, 0.01$~eV.  
Our mass range also compares favourably with the expected sensitivity  
$\sum_i (g_{\nu_i}/2)\,m_{\nu_i}\gwig\, 0.3$~eV of near-future global cosmological analyses involving 
also new CMB measurements~\cite{Croft:1999mm,Hu:1998mj}, as well as $m_\beta\gwig\, 0.75\div 1.8$~eV   
expected from neutrino observations from supernovae explosions~\cite{Beacom:2000ng,Arnaud:2002gt}.

From our fits, we could also determine the necessary ultrahigh energy neutrino flux at the 
resonance energy. Its prediction, in the context of the Z-burst scenario, turned out to be
of similar robustness as the prediction of the mass of the heaviest neutrino. It was found   
to be consistent with present upper limits and detectable in the near future by the already 
operating neutrino telescopes AMANDA and RICE, and by the Pierre Auger Observatory, presently
under construction.
A search at these facilities is the most promising and timely step in testing the Z-burst hypothesis.   
One does not have to wait for future projects such as EUSO~\cite{Catalano:2001mm} for this
critical investigation.  

The required neutrino fluxes are enormous. 
If such tremendous fluxes of ultrahigh energy neutrinos are indeed found, one has to deal with
the challenge to explain their origin. It is fair to say, that at the moment no convincing astrophysical
sources are known which meet the requirements for the Z-burst hypothesis, i.e. which  
have no or a negative cosmological evolution, 
accelerate protons at least up to $10^{23}$~eV, are opaque to primary
nucleons and emit secondary photons only in the sub-MeV region. 
Ideas to create the required neutrino fluxes from the decays of superheavy particles or topological defects
do not seem to be too appealing since they invoke further new physics beyond neutrino masses.

\begin{acknowledgments}
We thank S.~Barwick, 
O.~Biebel, S.~Bludman, W.~Buchm\"uller, P.~Di~Bari, M.~Kachelriess, K.~Mannheim, H.~Meyer,  
W.~Ochs, K.~Petrovay, S.~Sarkar, F.~Schrempp, D.~Semikoz, G.~Sigl, and E.~Waxman for useful 
discussions. 
We thank the OPAL collaboration for their unpublished
results on hadronic Z decays.
This work was partially supported by Hungarian Science Foundation
grants No. 
OTKA-\-T34980/\-T29803/\-T22929/\-M28413/\-OMFB-1548/\-OM-MU-708.
\end{acknowledgments}


\end{document}